\documentclass[preprint1]{aastex6}
\usepackage{graphicx}
\usepackage{rotating}
\usepackage{natbib}
\usepackage{aasref}
\usepackage{wrapfig}
\usepackage{placeins}
\usepackage{xcolor}

\begin{document}

\title{A curious ringlet that shares Prometheus' orbit \\ but precesses like the F ring}
\author{M.M. Hedman and B.J. Carter}
\affil{Physics Department, University of Idaho, Moscow ID 83844-0903}

\begin{abstract}
{Images obtained by the Cassini spacecraft of the region just beyond Saturn's main rings reveal a previously unreported narrow and dusty ringlet that has dynamical connections with both Saturn's small satellite Prometheus and the F ring. The radial position of this ringlet is observed to vary with time and longitude, indicating that it is eccentric with an eccentricity of 0.0012 and that its mean orbital radius varies between 139,300 km and 139,400 km. These mean radii are consistent with material trapped in a co-orbital 1:1 resonance with Prometheus. However, the apsidal precession rate of this ringlet is not that expected for material close to Prometheus' orbit (2.76$^\circ$/day). Instead, the ringlet appears to be precessing at the same rate as the F ring (2.70$^\circ$/day). This ringlet therefore appears to consist of material co-rotating with Prometheus whose apsidal precession rates have been modified by collisions with F-ring material. This ringlet may therefore provide new insights into how rings can maintain organized eccentric structures over a range of semi-major axes.}

\end{abstract}

\maketitle

\section{Introduction}

The region around the outer edge of Saturn's main rings corresponds to the Roche Limit for porous ice-rich objects \citep{Tiscareno13}, and so represents a dynamically rich area occupied by both rings and several small moons. Thus far, most of the work on this region has focused on the F ring and its interactions with the satellite Prometheus. The F ring is a narrow ring with a central strand between 10 and 100 km wide that exhibits a complex array of clumps, knots and fine scale structures \citep{Bosh97, Showalter04, Murray08}. The visible ring is dominated by dust-sized grains (i.e. less than 100 microns wide), but it contains a very narrow (probably less than 1 km wide) discontinuous core of larger particles that likely contain most of the F-ring's mass \citep{Cuzzi88, Esposito08, Beurle10, Hedman11, Vahidinia11, Meinke12, Attree12, French14}. The total mass of this ring is still uncertain, but is unlikely to exceed the mass of the nearby moon Prometheus \citep{Showalter92, Murray96}. This ring is also flanked on either side by additional strands of dusty material that can extend over several hundred kilometers in radius \citep{Murray97, Charnoz05}. One of the most puzzling aspects of this ring is that despite its complex and time-variable internal structure \citep{French12}, its overall shape is remarkably stable, with its main strand following the trajectory of a freely-precessing eccentric, inclined orbit with a semi-major axis of 140,221 km and an eccentricity of 0.00235 \citep{Albers12}. The coherent shape of the F ring is especially surprising because it lies just exterior to the small moon Prometheus, whose orbit has a semi-major axis of 139,380 km and an eccentricity of 0.0022 \citep{Spitale06, Jacobson08}. Since the pericenter of Prometheus' orbit drifts relative to that of the F ring, every 17 years the moon can get within 200 km of the F-ring's core. Indeed, Cassini has observed the moon producing a variety of structures within the F-ring's central strand \citep{Murray08}. It might at first appear that such disturbances would act to disperse the F ring and destroy its coherent structures, but recent work has suggested that interactions with Prometheus could help material in the ring coagulate into larger bodies via gravitational instabilities \citep{Beurle10} and maybe even confine these larger objects in semi-major axis via a complex interplay of resonances \citep{Cuzzi14}. However, despite these advances, many aspects of the F-ring's structure and dynamics remain obscure.

\begin{figure}
\centerline{\resizebox{6.5in}{!}{\includegraphics{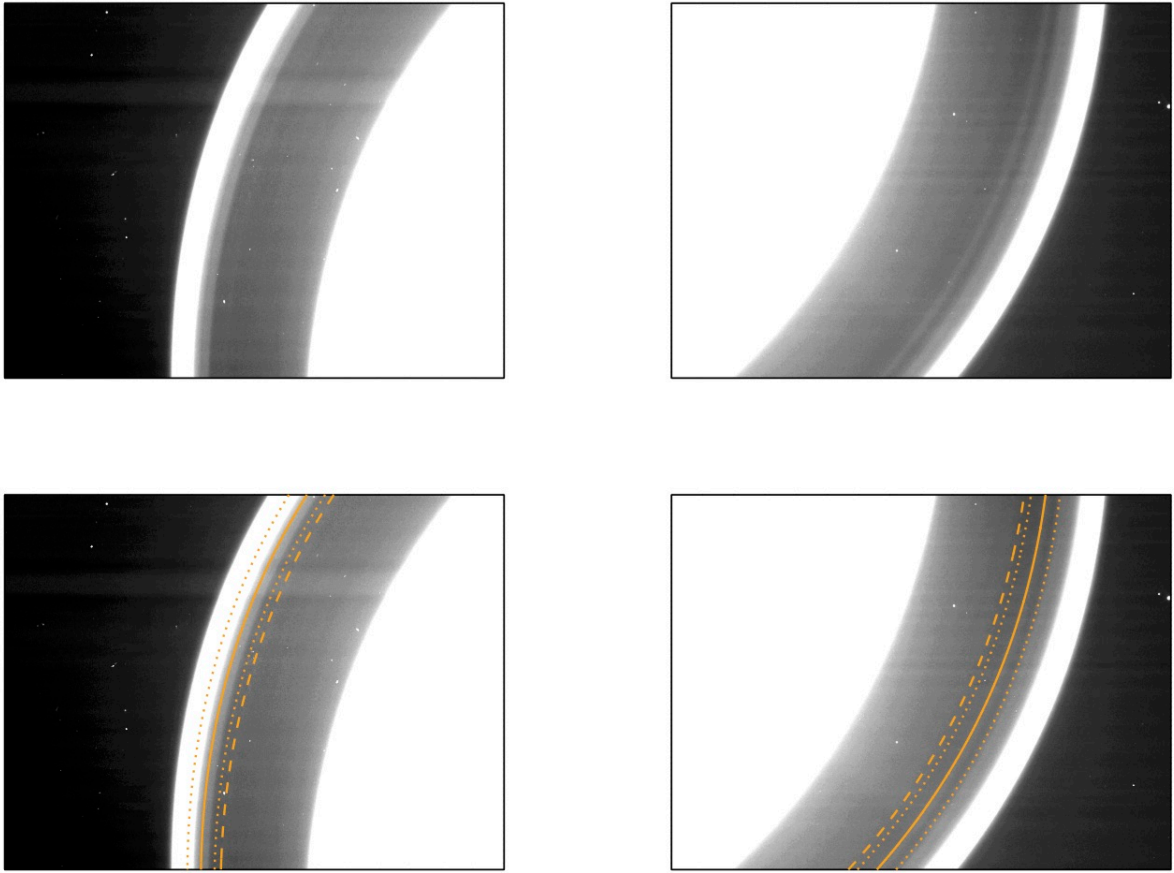}}}
\caption{Two images of the region between the A and F rings obtained on Day 300 of 2014 by the Narrow-Angle Camera onboard the Cassini Spacecraft (N1793077984 on left and N1793092620 on right). Both the A and F rings are overexposed in these images in order to show the material in between these rings. In both images there are bright and dark bands in the outer part of this region, near the F ring. The lower panels indicate the location of the ringlet R/2004 S2 (dashed line), Prometheus' semi-major axis (solid line) and the range of radii Prometheus moves through on its eccentric orbit (dotted lines). In both images, the brightness does decrease exterior to the nominal R/2004 S2 position, but there also appears to be a ringlet further out, whose peak brightness falls very close to Prometheus' semi-major axis. This ringlet appears as a distinct bright band in the right-hand image, when the F ring is close to its orbital apocenter, but also can be seen in the left-hand image as a ``shelf" on the F-ring's inner flank (close inspection reveals that there is a faint and narrow brightness minimum between the ringlet and the inner edge of the F ring in this image). }
\label{promringim}	
\end{figure}

New insights into the dynamics of this region can be obtained by examining the interactions between Prometheus and the other dusty rings in its vicinity. A broad sheet of dust extends inwards from the F ring across the entire Roche Division to the outer edge of the A ring \citep{Porco05}. Since Prometheus is embedded in this material, it should influence its structure. Indeed, images taken by the Cassini spacecraft when it first arrived at Saturn revealed a ringlet (designated R/2004 S2) at 138,900 km, just interior to Prometheus' orbit \citep{Porco04iau, Porco05, Winter05}. However, more recent images  reveal that this is not the only ringlet in this region. For example, Figure~\ref{promringim} shows two images of opposite side of Saturn's rings obtained by the Cassini spacecraft in late 2014. In these images the R/2004 S2 ringlet at 138,900 km is rather indistinct, and the most obvious ringlet in this region instead lies within the range of radii spanned by Prometheus' eccentric orbit. This ringlet is several hundred times fainter than the F ring and is sometimes difficult to discern against the  inner flank of that ring. However, it can be found in many Cassini images of this region obtained between 2006 and 2015, and so it appears to be a persistent structure. Furthermore, there are a few images where both this ringlet and the R/2004 S2 ringlet are visible, so the material found near Prometheus' orbit also appears to be distinct from that ringlet.  The most straightforward explanation for this feature is that it consists of material trapped in a 1:1 mean-motion resonance with Prometheus. Images of this material in the vicinity of Prometheus support this idea, because even though the shape and location of the ringlet are  different on either side of the moon, it does not contain structures like the F-ring's streamers that would be expected to arise in material not trapped in the resonance (see Figure~\ref{promimage} and Section~\ref{discussion} below). Other aspects of this ring's structure and dynamics are also consistent with this picture (see Section~\ref{discussion}). Hence for the remainder of this work we will refer to this feature informally as the ``Prometheus ringlet''.

This paper describes our initial investigations of the Prometheus ringlet, which has some interesting and unexpected properties that might provide new insights into the processes that enable the dust in and around the F ring to possess coherent, long-lived structures. First, we describe the data used in this analysis in Section~\ref{data}. Next, we take a qualitative look at these data in Section~\ref{first} in order to illustrate some interesting trends in these measurements. Then in Section~\ref{fits} we describe our techniques for extracting quantitative estimates of the ringlet's position at different times and locations and use this information to determine the orbital properties of this ringlet. These calculations reveal that the mean orbital radius of the ringlet particles is close to Prometheus' semi-major axis, but varies with longitude relative to that moon. The ringlet also has a non-zero orbital eccentricity that is significantly lower than the moon's. Most surprisingly, the pericenter of the Prometheus ringlet does not appear to precess around Saturn at the same rate as Prometheus' orbit, but instead maintains a fixed orientation relative to the F ring. Finally, Section~\ref{discussion} discusses how this anomalous precession rate could potentially arise due to collisions between particles in the Prometheus ringlet and F-ring material.

\begin{figure}
\centerline{\resizebox{6.5in}{!}{\includegraphics{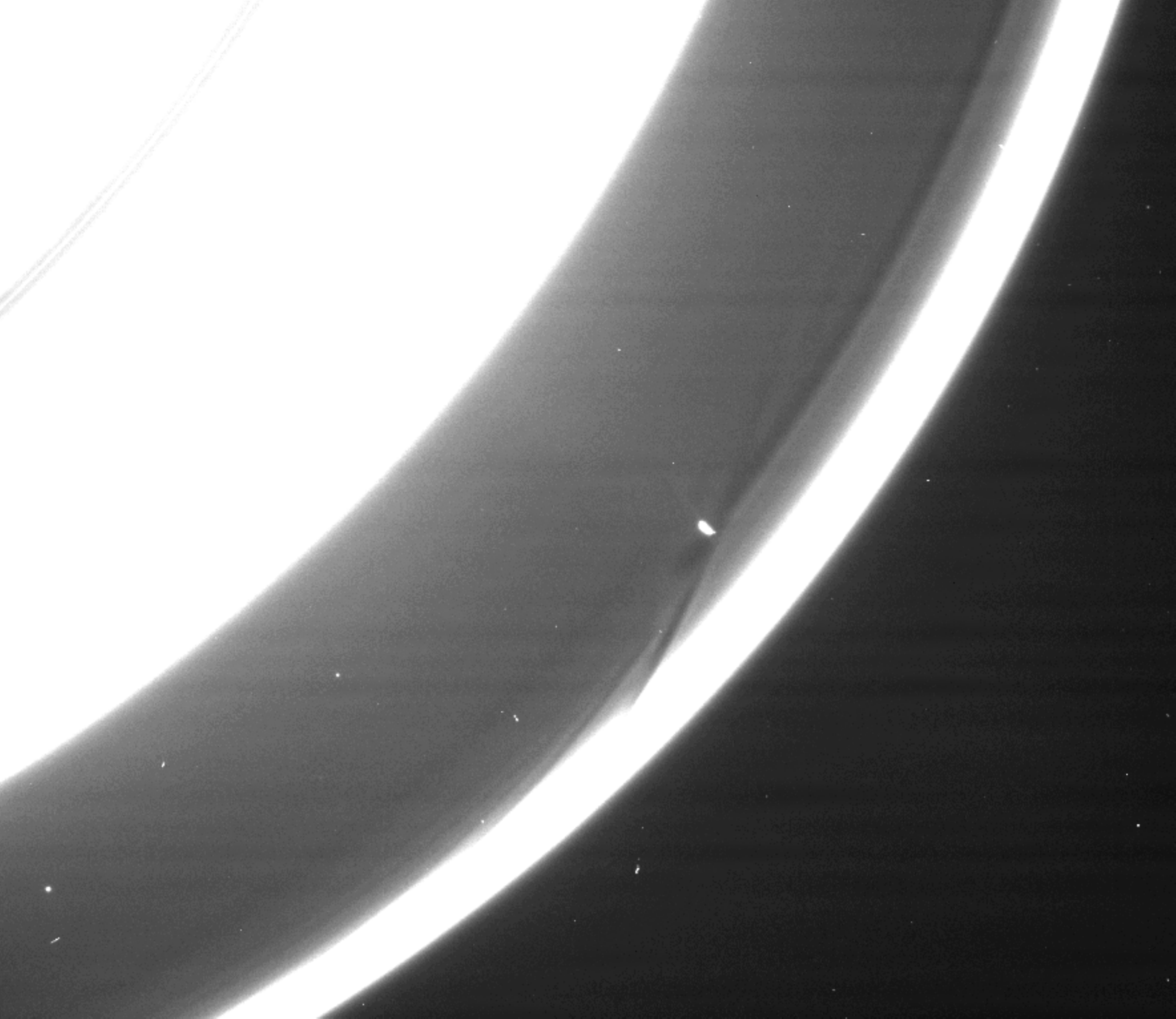}}}
\caption{An image of the Roche Division material in the vicinity of Prometheus  taken of Day 103 of 2014 (N1776123342). The A ring and F ring are overexposed to show the fainter dusty material. Both the moon and ring material are moving from the lower left to upper right in this image. Just behind and exterior to the moon are F-ring streamers \citep{Murray05}. The Prometheus ringlet can be seen most clearly ahead of Prometheus as the brightness maximum just outside the dark band. However, the ringlet can also seen as a fainter brightness maximum near the moon's orbit behind the moon and interior to the F-ring streamers. Note that while the structure of the ringlet is different on either side of the moon, it does not appear to contain periodic disturbances analogous to the F-ring streamers.}
\label{promimage}	
\end{figure}
	
\section{Observations} 
\label{data}

This analysis uses images obtained by the Narrow Angle Camera (NAC) of the Imaging Science Subsystem \citep{Porco04}. These data are all calibrated using the standard CISSCAL routines that apply flat-field corrections, remove dark currents and convert the measured data numbers into $I/F$, a standard measure of reflectance \citep{Porco04, West10}. Since this investigation focuses on the ringlet's structure, we will only consider images obtained through the camera's clear filters, which are both the most common images available and the ones with the highest signal-to-noise.

The Prometheus ringlet lies close to the F ring, and so is captured in many of the image sequences designed to monitor the F-ring's structure and evolution. However, since the Prometheus ringlet is several hundred times fainter than the F ring, it is not easy to see in many F-ring images. We therefore focused our attention on sequences which repeatedly imaged the F ring at phase angles above 125$^\circ$. Since the ringlet is strongly forward-scattering, it is easiest to see at these high phase angles. We identified eleven observation in which the ringlet could be clearly detected and its position reliably estimated using our algorithms (see below). For most of these observations, the camera stared at either one or two locations in the ring and watched material as it rotated through the field of view. However, for one observation (designated FMOVIE199, see below) the spacecraft did not stare at fixed locations, but instead tracked a particular point in the ring as it moved around the planet. Table~\ref{dattab} summarizes the properties of the relevant observations. Note that five of the  observations contain two distinct image sets where the camera was pointed at two different locations in the rings. Each of these image sets is given a separate entry in the table. For convenience's sake, we designate each image set within an observation with a name composed of the relevant observation's name (either FMOVIE or FRSTRCHAN) followed by the so-called ``Rev" number, which corresponds to Cassini's orbit around Saturn. If needed, we also use a letter to distinguish between the two image sets that are part of the same observation (e.g. FMOVIE209a and FMOVIE209b).



Each image in all these  observations was geometrically navigated using the relevant SPICE kernels \citep{acton96} listed in Table~\ref{kernels}, and the nominal camera pointing was refined based on the positions of known stars in the field of view (images without sufficient stars were removed from further consideration). After each image was navigated, the brightness data in each image were averaged over all longitudes to produce a high signal-to-noise profile of the ring's brightness as a function of radius. While the ringlet is visible in many of these profiles, the background trends associated with the F-ring's inner flank complicate efforts to visualize and quantify this feature. Hence we remove a smooth background model in order to better isolate the signal from the desired ringlet. The model of the background trends for each profile was computed by fitting the logarithm of the data between 138,000 km and 140,000 km to a fourth-order polynomial (excluding any data where the brightness is more than 5 times its minimum value in that range). Initially, we excluded the region containing the Prometheus ringlet from these fits because we thought the background model would remove some signal from the ringlet, but after some experimentation we found that the ringlet signal is narrow enough for this not to be a problem (The ringlet is a relatively small and compact fluctuation, and did not affect the fit to the broad and steep background curve of the inner F ring). These background-subtracted profiles form the basis of the following analysis.

\begin{sidewaystable}[tbp]
\caption{Images considered for this study}
\label{dattab}
\centerline{\resizebox{9in}{!}{\begin{tabular}{|c|c|c|c|c|c|c|c|c|c|}\hline
Observation ID $^a$ & Files & Image Set$^b$ & Observation & Duration & Observation Time & Phase & Emission & Inertial & Longitude Relative \\  & & & Date & (hours) & (ET Seconds) &  Angle & Angle & Longitude$^c$ & to Prometheus$^c$ \\ \hline  
 \textcolor{black}{ISS\_029RF\_FMOVIE001\_VIMS} & N1538168640-N1538218132(92) & FMOVIE029a & 2006-272 & 13.7 & 212747594.-212797086. & 158.4-161.8 &  57.5- 59.1 &  265.8- 269.8 & -148.2- 184.3 \\
 \textcolor{black}{ISS\_029RF\_FMOVIE001\_VIMS} & N1538269441-N1538300071(54) & FMOVIE029b & 2006-273 &  8.5 & 212848395.-212879024. & 158.9-161.0 &  60.6- 61.5 &   83.2-  85.5 & -169.0-  36.5 \\
 \textcolor{blue}{ISS\_031RF\_FMOVIE001\_VIMS} & N1541012989-N1541062380(110) & FMOVIE031 & 2006-305 & 13.7 & 215591925.-215641316. & 156.3-160.3 &  51.9- 55.2 &   94.7-  99.3 & -211.2- 119.8 \\
 \textcolor{cyan}{ISS\_032RF\_FMOVIE001\_VIMS} & N1542047596-N1542156952(128) & FMOVIE032 & 2006-317 & 30.4 & 216626526.-216735881. & 155.9-163.9 &  51.8- 59.0 &   94.4- 103.8 &  -87.2- 286.6 \\
 \textcolor{green}{ISS\_173RF\_FMOVIE001\_PRIME} & N1729024626-N1729053296(53) & FMOVIE173a & 2012-289 &  8.0 & 403602297.-403630966. & 137.5-142.1 & 123.9-124.2 &  140.1- 109.5 &  -18.3- 207.0 \\
 \textcolor{green}{ISS\_173RF\_FMOVIE001\_PRIME} & N1729053606-N1729081745(52) & FMOVIE173b & 2012-290 &  7.8 & 403631277.-403659415. & 153.4-160.6 & 118.5-123.3 &  335.3- 305.4 &  -16.1- 205.4 \\
 \textcolor{green!50!black}{ISS\_174RF\_FRSTRCHAN001\_PRIME} & N1731106419-N1731132308(62) & FRSTRCHAN174a & 2012-314 &  7.2 & 405684077.-405709965. & 150.4-156.0 & 121.2-123.6 &  320.2- 320.1 &  100.8- 276.4 \\
 \textcolor{green!50!black}{ISS\_174RF\_FRSTRCHAN001\_PRIME} & N1731132699-N1731158588(60) & FRSTRCHAN174b & 2012-314 &  7.2 & 405710356.-405736245. & 143.9-149.5 & 119.3-121.6 &  137.0- 136.9 &   98.6- 275.0 \\
 \textcolor{red}{ISS\_196RF\_FMOVIE006\_PRIME} & N1756377239-N1756428571(120) & FMOVIE196 & 2013-240 & 14.3 & 430954736.-431006068. & 145.3-150.1 & 129.5-132.7 &  300.1- 299.6 & -145.1- 204.3 \\
 \textcolor{magenta}{ISS\_199RF\_FMOVIE002\_PRIME} & N1765017544-N1765070521(117) & FMOVIE199 & 2013-340 & 14.7 & 439594986.-439647963. & 135.9-147.2 & 126.6-132.4 & -195.6- 174.4 &   28.2-  18.3 \\
 \textcolor{red!40!brown}{ISS\_201RF\_FMOVIE001\_VIMS} & N1770315948-N1770367142(157) & FMOVIE201 & 2014-037 & 14.2 & 444893357.-444944550. & 125.8-142.8 & 131.0-137.6 &  304.8- 326.8 & -181.7- 144.2 \\
 \textcolor{violet}{ISS\_203RF\_FMOVIE001\_PRIME} & N1776077815-N1776106170(66) & FMOVIE203a & 2014-103 &  7.9 & 450655187.-450683542. & 136.5-139.8 &  72.1- 74.5 &  201.9- 201.8 &  -76.1- 116.3 \\
 \textcolor{violet}{ISS\_203RF\_FMOVIE001\_PRIME} & N1776106495-N1776134850(70) & FMOVIE203b & 2014-103 &  7.9 & 450683867.-450712222. & 129.0-132.0 &  70.3- 72.2 &   35.4-  35.4 &  -77.8- 115.3 \\
 \textcolor{gray}{ISS\_208RF\_FMOVIE001\_PRIME} & N1789913768-N1789932276(90) & FMOVIE208 & 2014-263 &  5.1 & 464491052.-464509560. & 131.9-142.4 & 126.5-132.6 &   52.0-  51.9 &  114.7- 240.1 \\
 \textcolor{orange}{ISS\_209RF\_FMOVIE001\_PRIME} & N1793068408-N1793089959(46) & FMOVIE209a & 2014-300 &  6.0 & 467645672.-467667223. & 132.8-134.2 &  71.2- 72.0 &  214.9- 214.8 &   54.0- 200.6 \\
 \textcolor{orange}{ISS\_209RF\_FMOVIE001\_PRIME} & N1793090708-N1793110822(43) & FMOVIE209b & 2014-300 &  5.6 & 467667972.-467688085. & 127.9-129.1 &  70.5- 71.2 &   40.0-  40.1 &   97.2- 234.0 \\ \hline
\end{tabular}}}

\medskip

$^a$ This is the standard designation for the observation as planned by the Cassini project. The color of this label corresponds to the colors used for the data points in Figures~\ref{fitalldat}-~\ref{fitlondat2}.

$^b$ This is a shorthand label for each sequence of images, which includes the Cassini Rev number (orbit around Saturn) and a letter designation if there are more than one set of images staring at a particular inertial longitude in a given observation.

$^c$ The longitude ranges shown here are signed quantities to clarify the range of longitudes covered in the image sequence. Most sequences cover a narrow range of inertial longitudes, but the FMOVIE199 observation spans nearly all inertial longitudes. For the longitudes relative to Prometheus, longitude ranges that span zero mean that the observed region includes material in the vicinity of that moon (compare with Figure~\ref{fmovieim}).

\end{sidewaystable}

\begin{table}
\caption{SPICE kernels used in this investigation}
\label{kernels}
\centerline{\begin{tabular}{c} 
naif0010.tls \\
cas00167.tsc \\
pck00010.tpc \\
cpck10Feb2010.tpc \\
cpck\_rock\_25Aug2008\_merged.tpc \\
150720AP\_RE\_90165\_18018.bsp \\
120711CP\_IRRE\_00256\_25017.bsp \\
061108R\_SCPSE\_06260\_06276.bsp \\
061129RB\_SCPSE\_06292\_06308.bsp \\
061213R\_SCPSE\_06308\_06318.bsp \\
070109R\_SCPSE\_06318\_06332.bsp \\
121204R\_SCPSE\_12257\_12304.bsp \\
130318R\_SCPSE\_12304\_12328.bsp \\
131024R\_SCPSE\_13200\_13241.bsp \\
131105R\_SCPSE\_13241\_13273.bsp \\
140409R\_SCPSE\_14025\_14051.bsp \\
140730R\_SCPSE\_14051\_14083.bsp \\
140907R\_SCPSE\_14083\_14118.bsp \\
150122R\_SCPSE\_14251\_14283.bsp \\
150304R\_SCPSE\_14283\_14327.bsp \\    
 \end{tabular}}
\end{table}

\begin{figure}
\resizebox{6.5in}{!}{\includegraphics{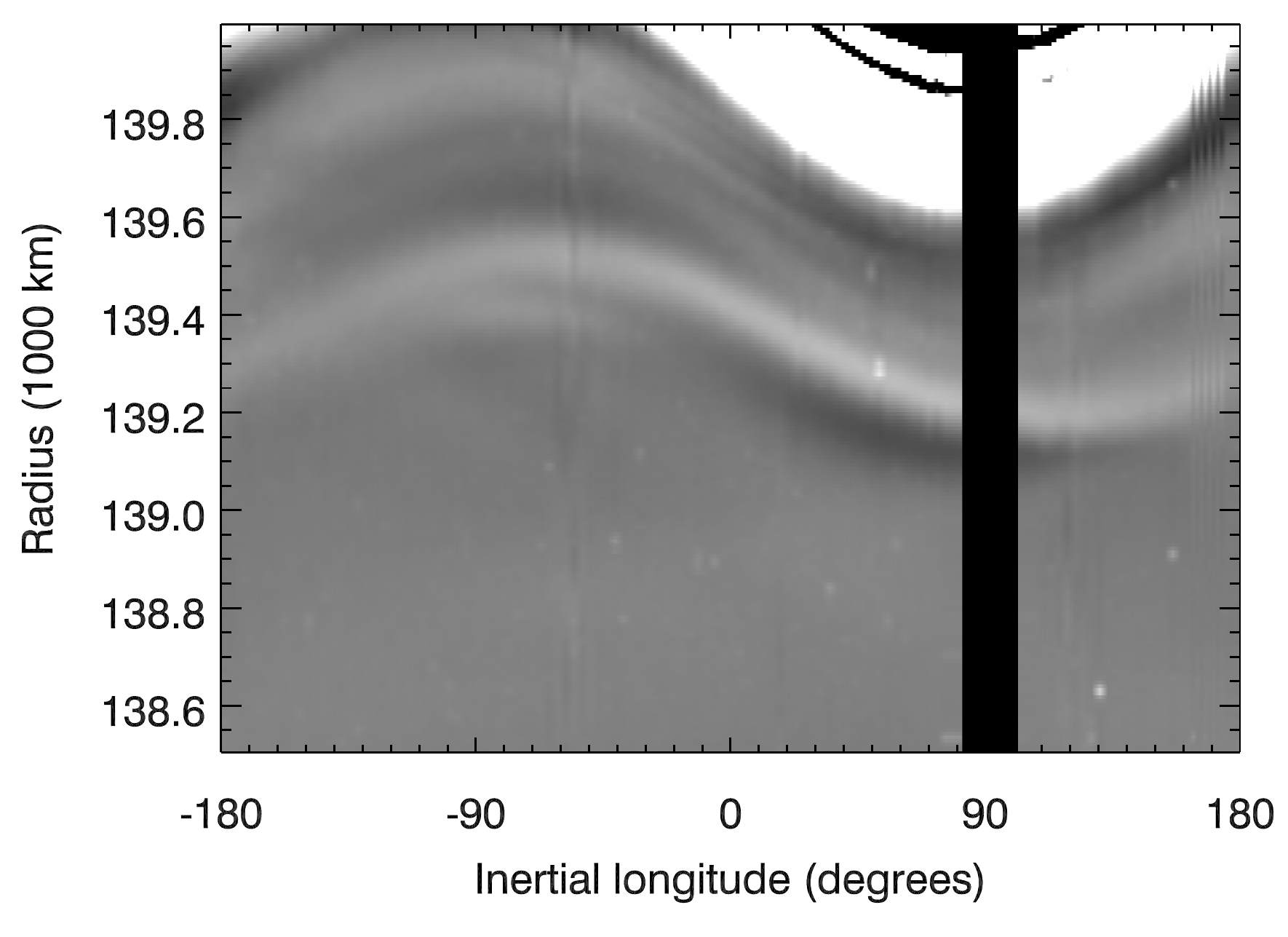}}
\caption{Map of the rings' (background-subtracted) brightness at a particular co-rotating longitude (about 20$^\circ$ in front of Prometheus) as a function of radius and inertial longitude derived from the FMOVIE199 sequence. Note the overexposed feature near the top of the image is the F ring,while the Prometheus ringlet is visible between 139,100 km and 139,600 km. The radial location of the ringlet varies with inertial longitude in a manner consistent with an eccentric ringlet.}
\label{fmovie199im}
\end{figure} 

\begin{sidewaysfigure}
\resizebox{3.in}{!}{\includegraphics{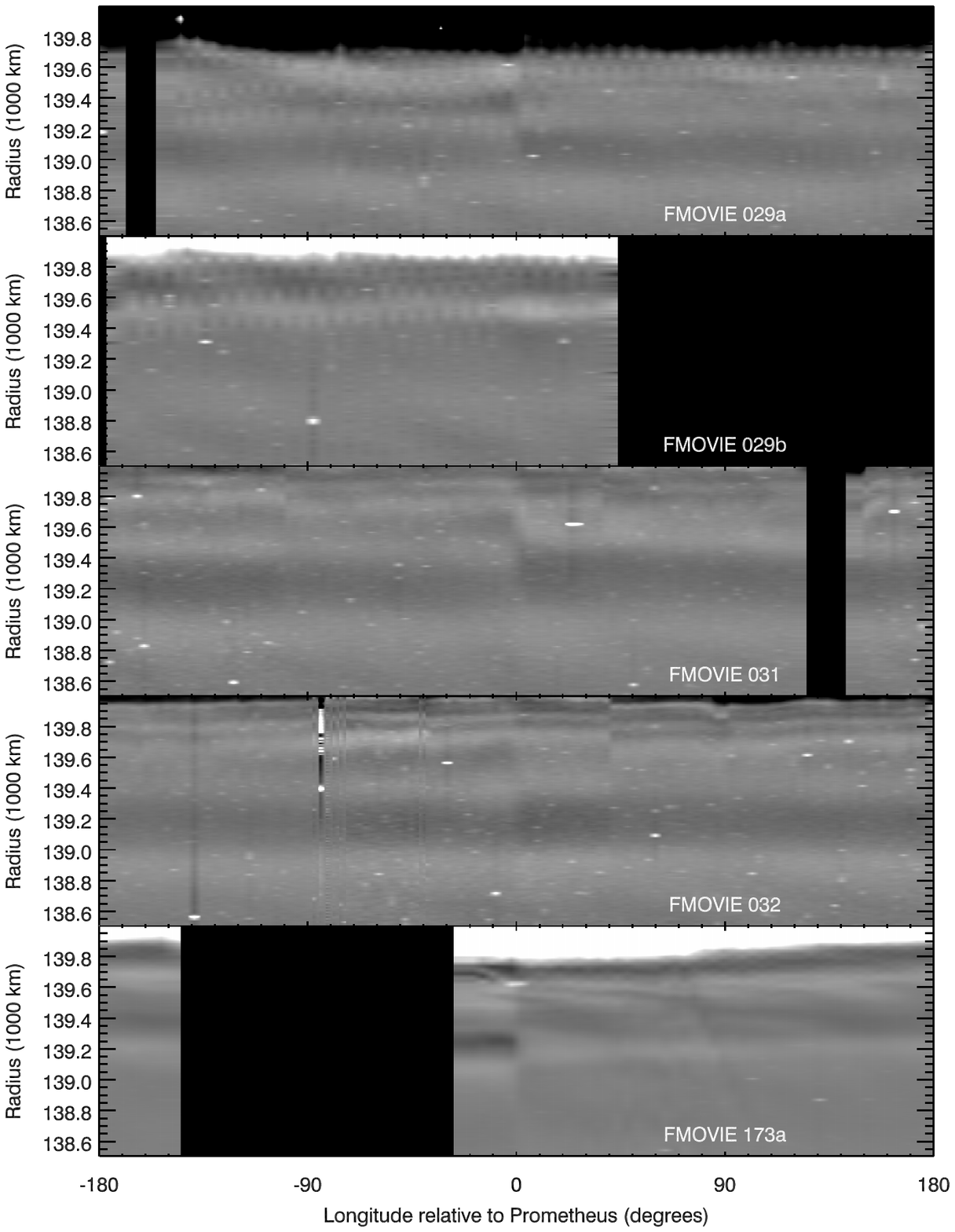}}
\resizebox{3.in}{!}{\includegraphics{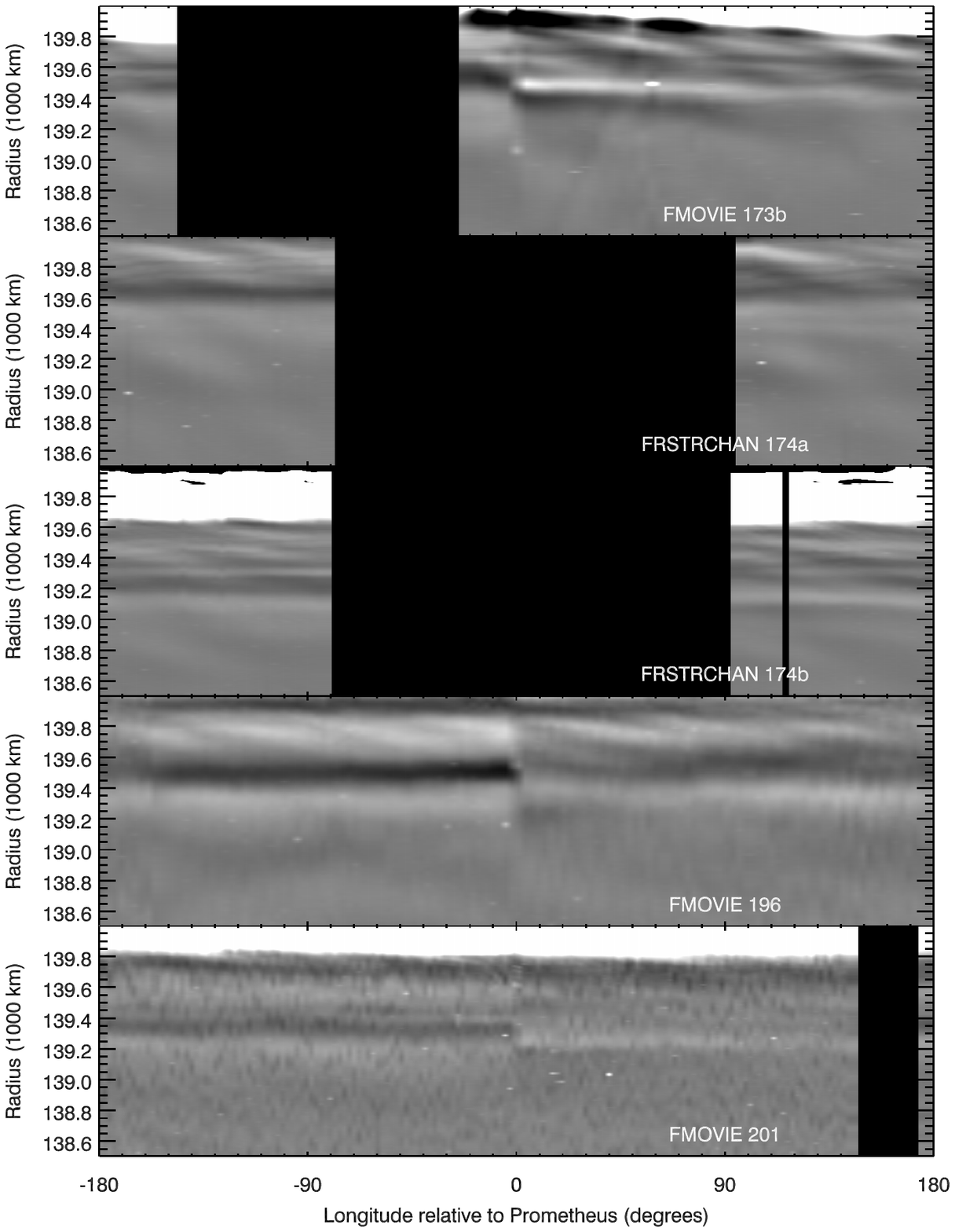}}
\resizebox{3.in}{!}{\includegraphics{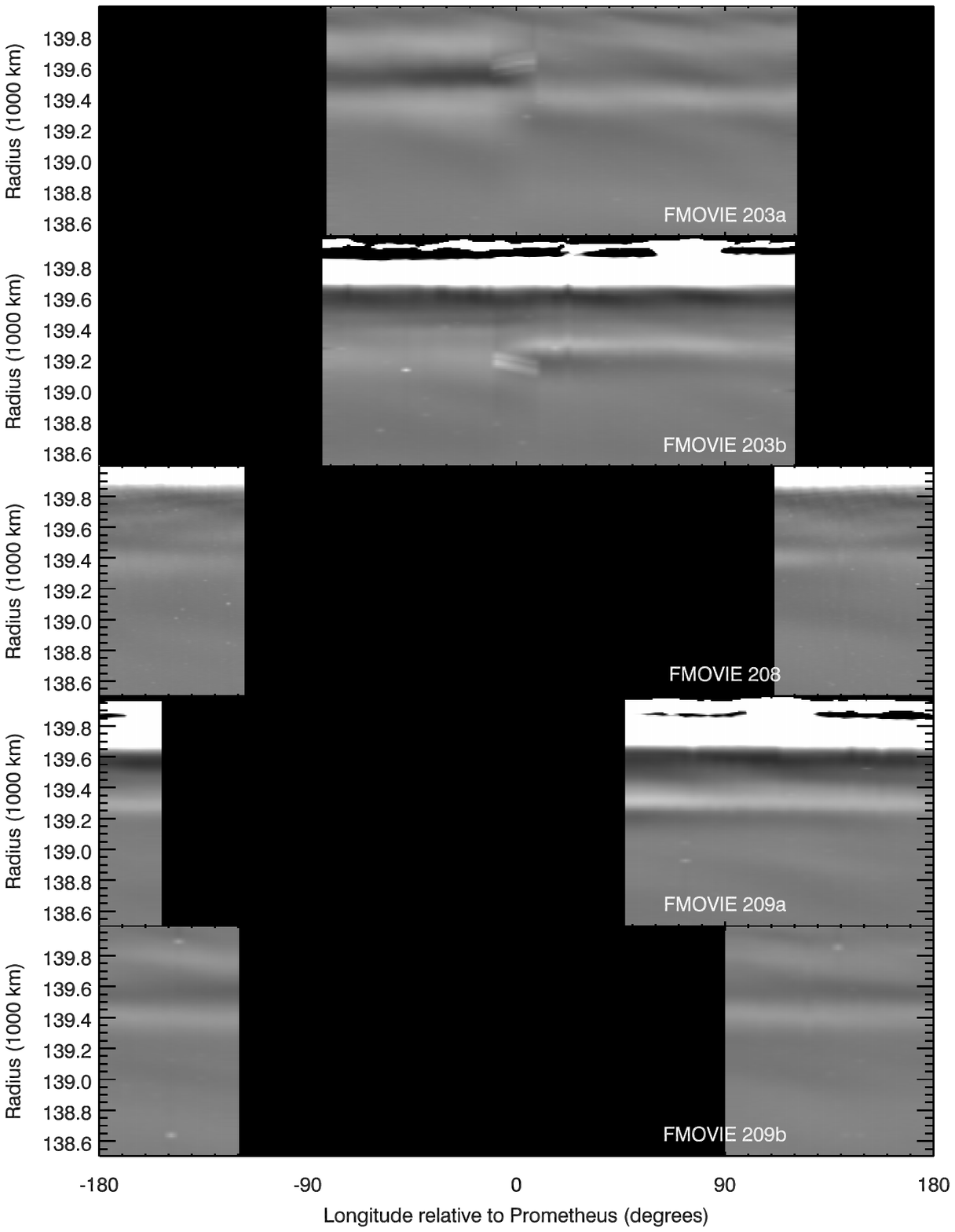}}
\caption{Maps of the Prometheus ringlet derived from the various observing sequences where the spacecraft stared at a single inertial longitude. Each map shows the rings' (background-subtracted) brightness at the relevant inertial longitude as a function of radius and co-rotating longitude relative to Prometheus (Blank areas are regions not covered in the particular observation). While the ringlet is easier to see in the later observations, it is detectable in all the data sets.}
\label{fmovieim}
\end{sidewaysfigure}

\pagebreak

\section{Overall structure of the ringlet}
\label{first}

One way to visualize the data encoded in the brightness profiles is by assembling them into maps of the ring's background-subtracted brightness as a function of radius and the appropriate longitudinal coordinate sampled by the various images. Note that there are two different longitudes that are relevant to this analysis. One is the ``inertial longitude", which is measured relative to a fixed direction is space (corresponding to the ascending node of the ringplane in the J2000 reference system), while the other is a ``co-rotating longitude''. Given the ringlet's close association with Prometheus, the co-rotating longitude is computed as the difference between the observed inertial longitude $\lambda$ and the longitude of Prometheus $\lambda_P$ derived from the SPICE kernels (see Table~\ref{kernels}). 

Figure~\ref{fmovie199im} shows a map of the ring's brightness at a single co-rotating longitude (about 20$^\circ$ in front of Prometheus) versus radius and inertial longitude derived from the FMOVIE199 sequence. The ringlet's radial location $r$ varies with inertial longitude $\lambda$ in a manner consistent with an eccentric ringlet:
\begin{equation}
r=a+ae\cos(\lambda-\varpi)
\end{equation}
where $a$, $e$ and $\varpi$ are the ringlet's mean orbital radius, eccentricity and pericenter location, respectively. Note that many dusty narrow rings (including the F ring) exhibit non-zero eccentricities, and so it is not unreasonable for the Prometheus ringlet to be eccentric. This observation also suggests that the ringlet could be found at any radius between 139,100 and 139,600 km.

Figure~\ref{fmovieim} shows maps of the rings derived from the other fifteen image sets. Each map displays the background-subtracted brightness of the rings at the observed inertial longitude as functions of radius and co-rotating longitude relative the Prometheus. In all these maps a faint ringlet can be seen somewhere between 139,100 and 139,600 km. The position of the ringlet most likely varies among the different image sets because the camera observed the ringlet at different longitudes and times, and hence at different positions relative to the ringlet's pericenter. However, these data also indicate that the exact location of the ringlet also varies with longitude relative to Prometheus. These trends can be seen most clearly in the FMOVIE196, 201 and 203 image sets, where the ringlet appears to be at larger radii just in front of Prometheus than it is just behind that moon. Indeed, it appears that the radial location of the ringlet shifts steadily inwards with increasing longitude relative to Prometheus. 

A similar trend in a ringlet's position with co-rotating longitude has been observed in the dusty ringlets that lie within the Encke Gap in Saturn's outer A ring. In particular, the Central Encke Gap ringlet, which shares its orbit with the small moon Pan, contains particles whose mean orbital radii steadily increase with azimuthal distance behind that moon up to a clump-rich region where the particles' mean orbital radius suddenly returns to Pan's semi-major axis \citep{Hedman13}. This outward trend was interpreted as the result of material being ejected from Pan at low velocities and drifting outwards under the influence of drag forces, which cause the dust particles to drift outwards and backwards relative to the moon. The trends observed here may have a similar origin, as will be discussed in more detail below.

\section{Fitting the ringlet's position and shape}
\label{fits}

\begin{figure}
\resizebox{6.5in}{!}{\includegraphics{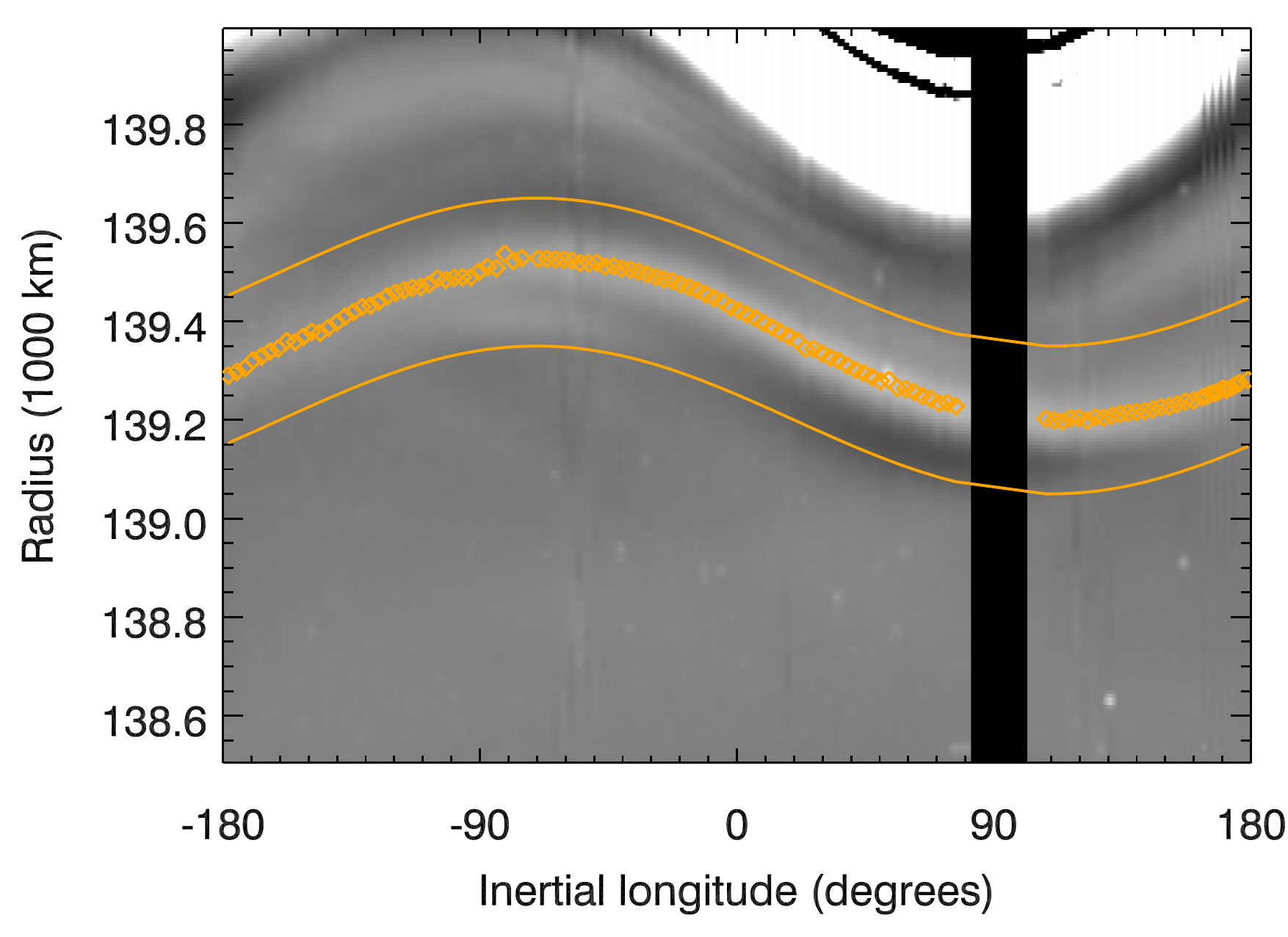}}
\caption{The estimated location of the Prometheus ringlet in the FMOVIE199 observation. The background image is the same map of the ring's background-subtracted brightness shown in Figure~\ref{fmovie199im}. The two lines show the radial region that was fit to a model of the ringlet's brightness, and the diamonds show the estimated ringlet position from the fit.}
\label{fmovie199fit}
\end{figure} 

\begin{sidewaysfigure}
\resizebox{3.in}{!}{\includegraphics{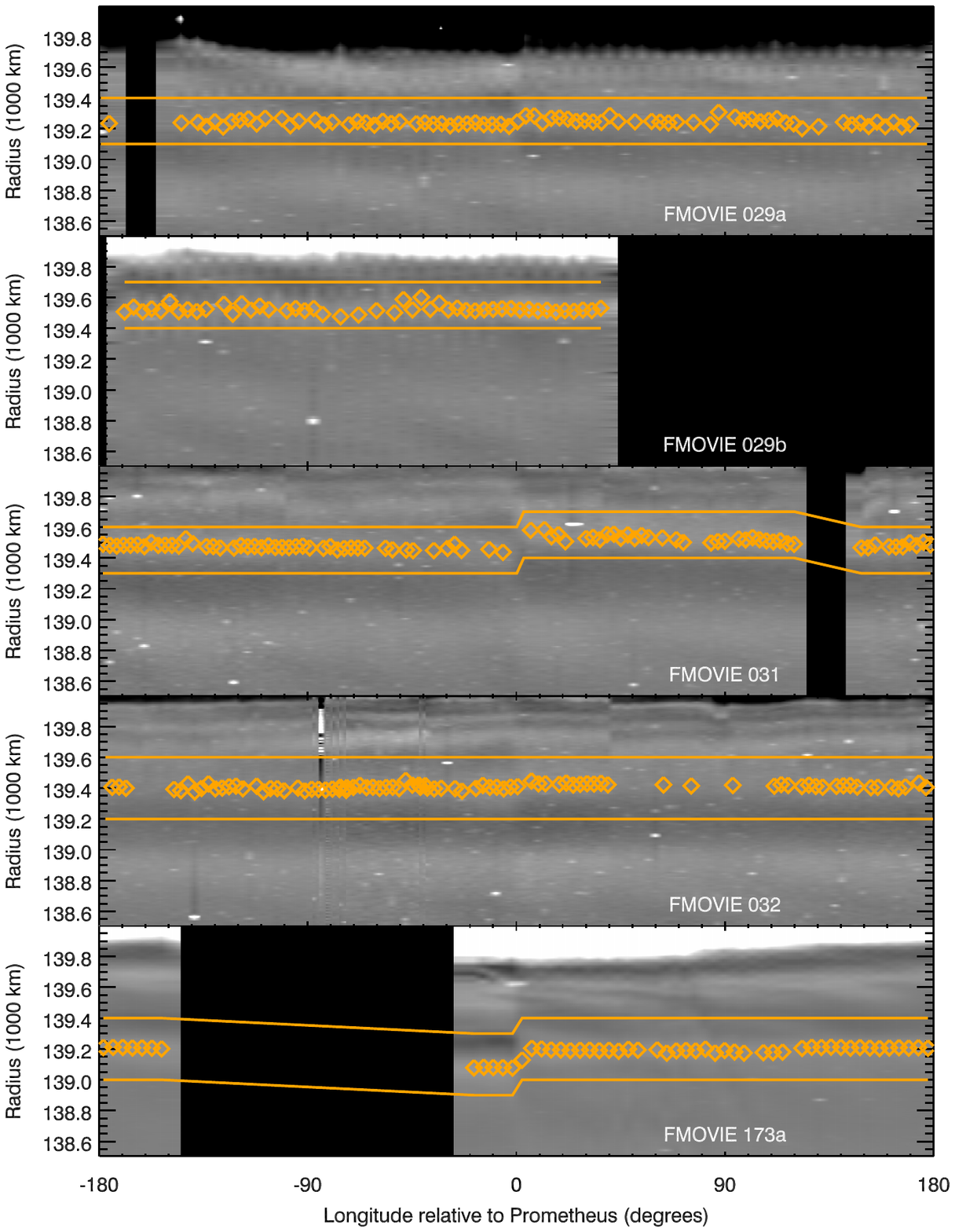}}
\resizebox{3.in}{!}{\includegraphics{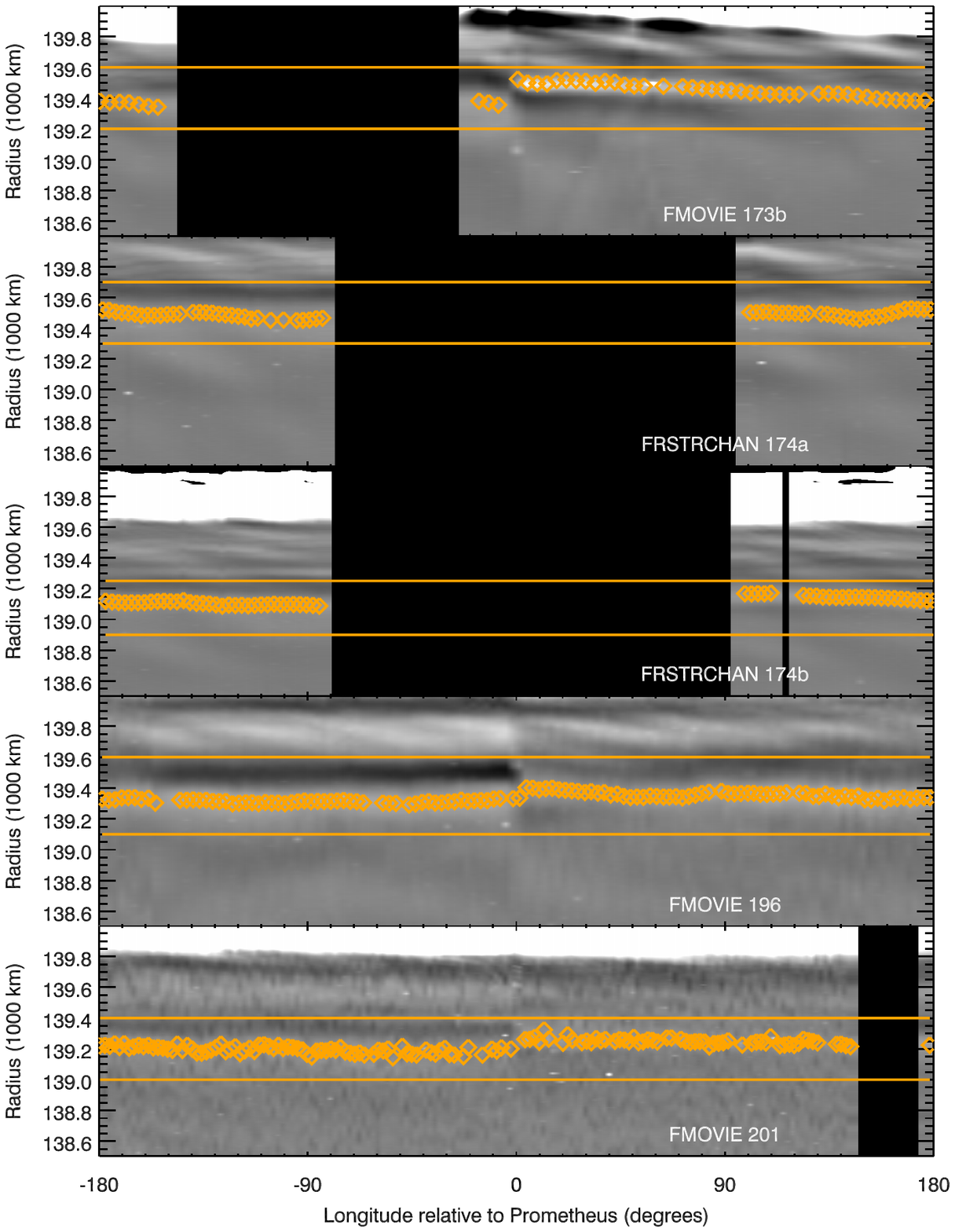}}
\resizebox{3.in}{!}{\includegraphics{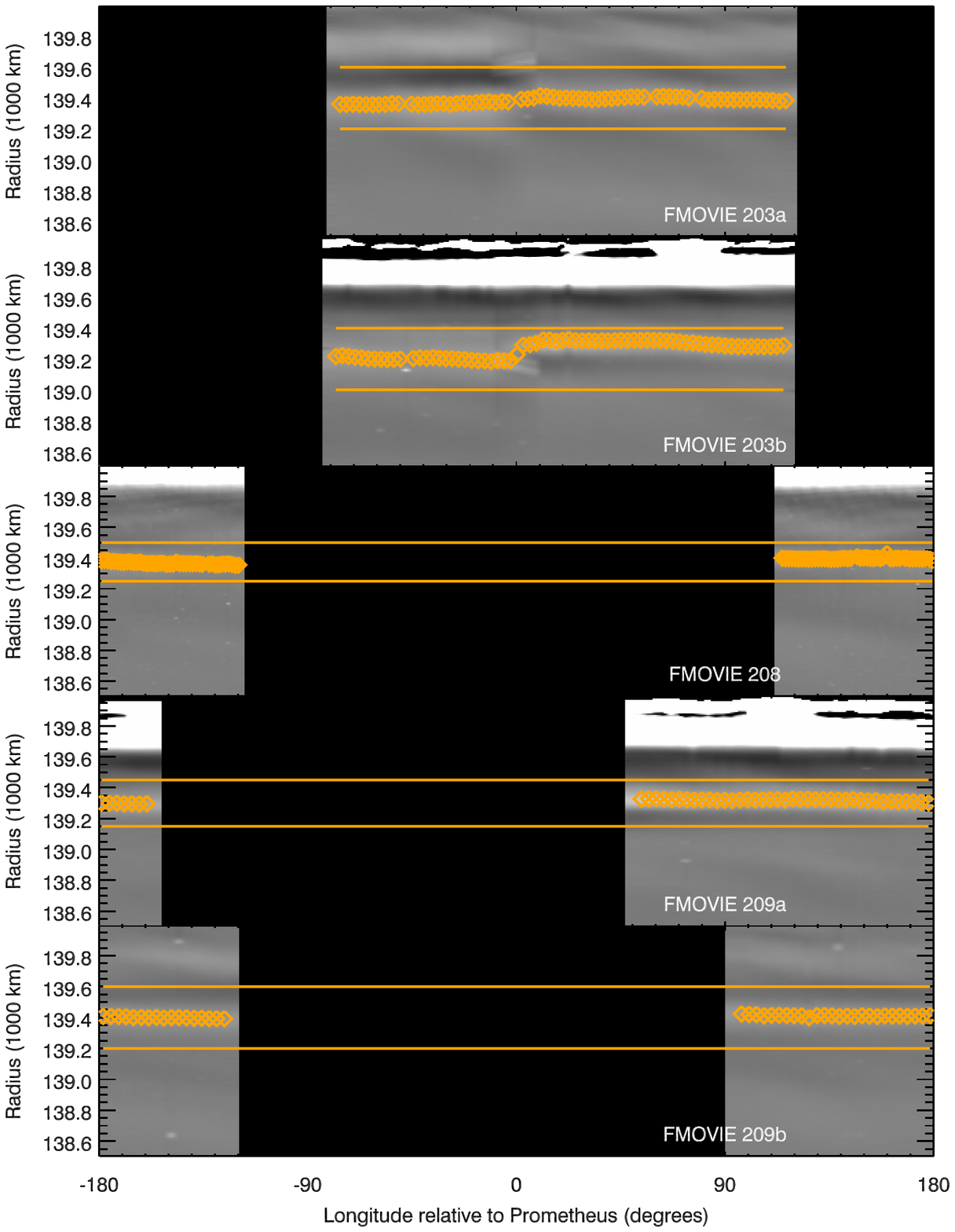}}
\caption{The estimated location of the Prometheus ringlet in the observations where the camera stared at a fixed inertial longitude. In each panel, the background image is the same map of the ring's background-subtracted brightness shown in Figure~\ref{fmovieim}. The two lines show the radial region that was fit to a model of the ringlet's brightness, and the diamonds show the estimated ringlet position from the fit.}
\label{fmoviefit}
\end{sidewaysfigure}

In order to explore these variations in the ringlets' radial position more quantitatively, we fit the background-subtracted brightness profiles in the vicinity of the ringlet's position to a Lorentzian peak plus linear background using the {\tt mpfitpeak} routine in IDL \citep{Markwardt09}. The exact range of radii fit in each profile was chosen to include the entire ringlet and to exclude any residual F-ring structures (these ranges are illustrated by the solid lines in Figures~\ref{fmovie199fit} and~\ref{fmoviefit}). For a few profiles, the fit fails to find the appropriate peak and instead gives the position of a star or other structure in the profile, but these bad fits are easily identified as those where the peak width is less than 50 km or more than 200 km, or where the fit gives a dip instead of a peak. After excluding those fits, we have 1202 position estimates from the various observations, which are shown as diamonds in Figures~\ref{fmovie199fit} and~\ref{fmoviefit}, and are provided in tabular form in supplemental online information. All of the selected position estimates are consistent with the apparent location of the ringlet in the relevant map. 

\begin{figure}
\resizebox{6.5in}{!}{\includegraphics{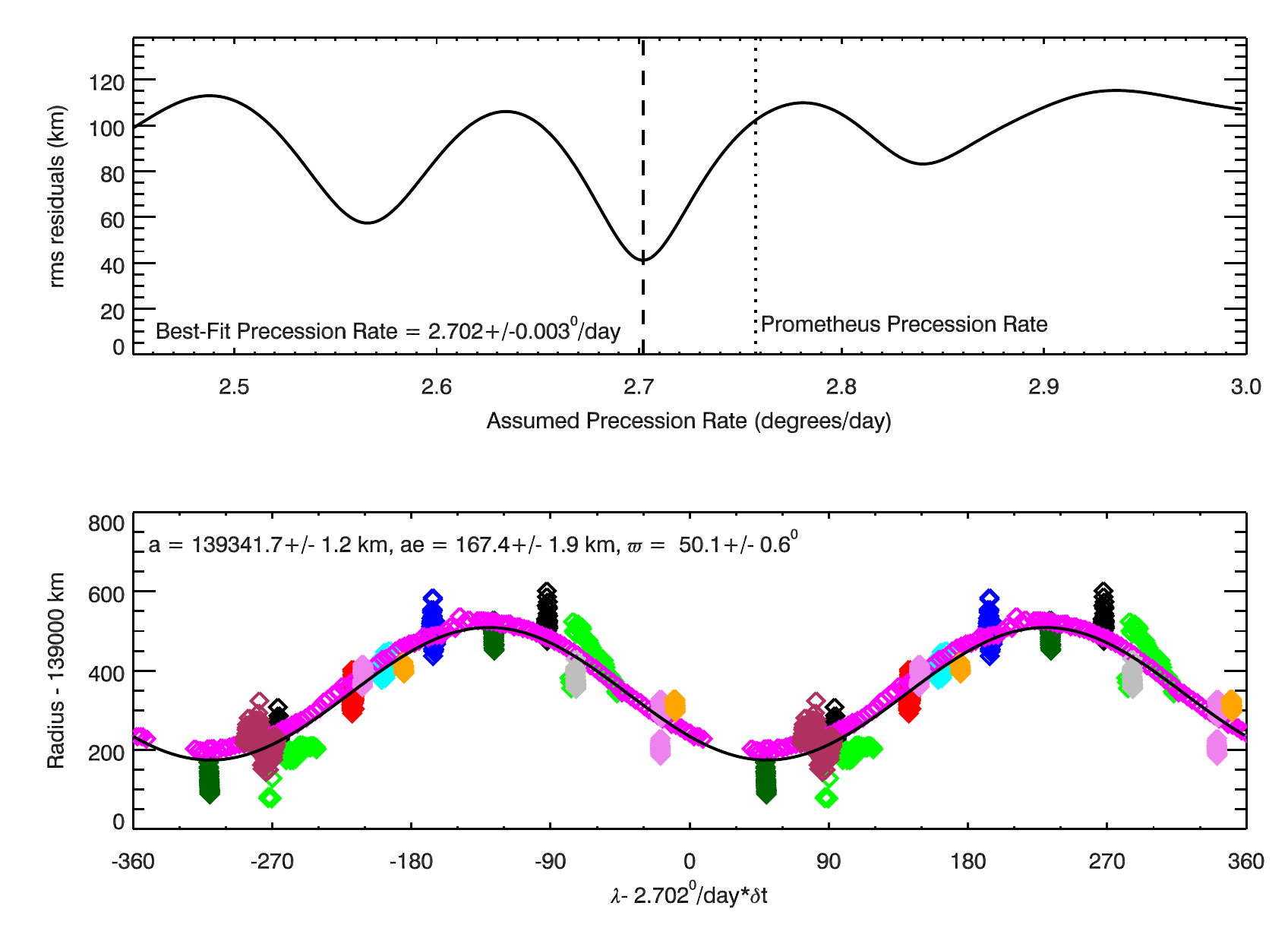}}
\caption{Fitting the position of the Prometheus ringlet to a precessing ringlet model. The top panel shows the $rms$ residuals of the fit as a function of the assumed precession rate. The dashed vertical line corresponds to the best-fit precession rate, while the dotted line is the precession rate of Prometheus from \citep{Spitale06, Jacobson08}. Note the best-fit precession rate is significantly slower than the expected precession rate for Prometheus. The bottom panel shows the observed radial position of the ringlet as a function of true anomaly for the best-fit precession rate at the J2000 epoch time, with different colors corresponding to different observations (see Table~\ref{dattab}).}
\label{fitalldat}
\end{figure} 

\begin{figure}
\resizebox{6.5in}{!}{\includegraphics{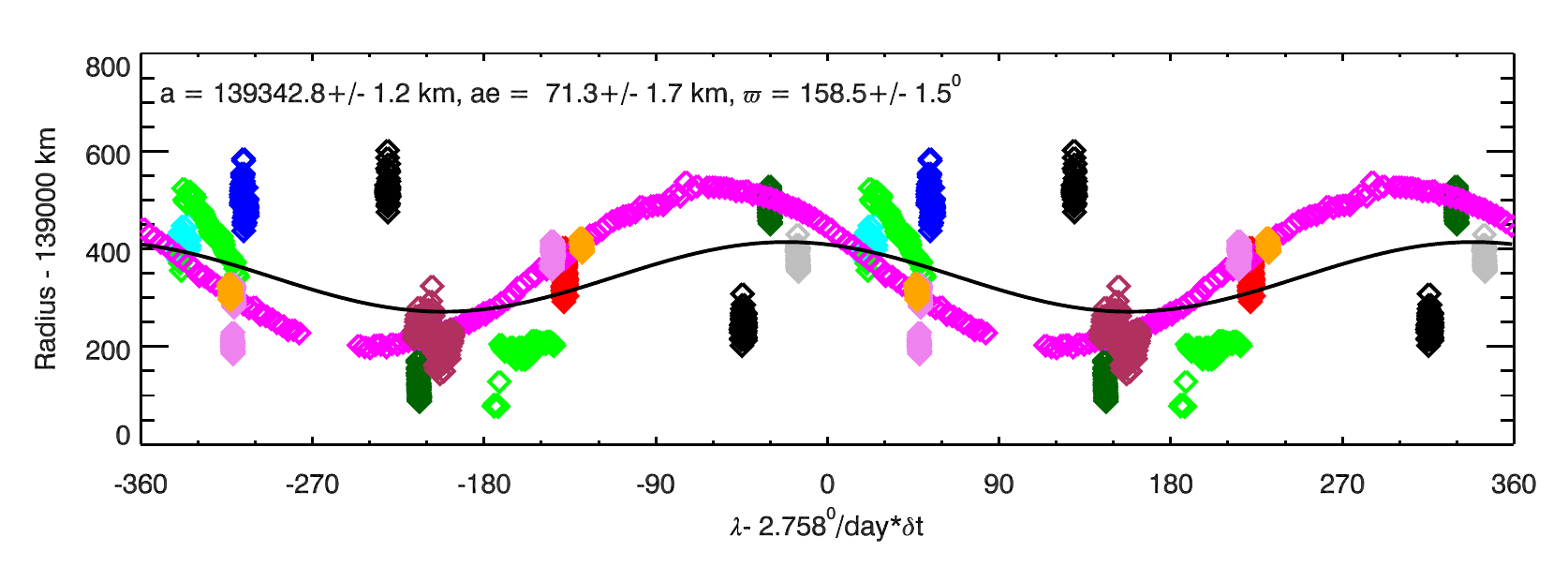}}
\caption{The observed radial position of the ringlet relative to a model where we have forced the ringlet's precession rate to match that of Prometheus' orbit (2.758$^\circ$/day). Note that the data are much less well organized in this case.}
\label{fitpromdat}
\end{figure} 

These position estimates, together with the corresponding longitudes and observation times, can then be used to constrain the shape of this ringlet. Since the variations in the ringlets' position with inertial longitude in Figure~\ref{fmovie199fit} are significantly larger than the trends with co-rotating longitude shown in Figure~\ref{fmoviefit}, it makes sense to first fit the overall eccentric shape of the ringlet and then consider the trends with co-rotating longitude relative to Prometheus. We therefore begin by fitting the radii $r$ as functions of the inertial longitude $\lambda$ and time $t$ (relative to J2000 epoch) to a model of a uniformly precessing ringlet:
\begin{equation}
r=a-ae\cos\left[\lambda-(\varpi_0+\dot{\varpi}t)\right]
\end{equation}
where $a$, $e$, $\varpi_0$ and $\dot{\varpi}$ are all constants that correspond to the ringlet's mean orbital radius (or effective semi-major axis), eccentricity, pericenter location at the epoch time (taken to be the J2000 epoch time here) and apsidal precession rate. We fit for these parameters using a two-step procedure. Assuming the accepted gravity model of Saturn \citep{Jacobson06,Jacobson08}, the precession rate near Prometheus should be 2.758$^\circ$/day. Therefore, we consider a range of precession rates between 2.5$^\circ$/day and 3.0$^\circ$/day with a step size of 0.001$^\circ$/day. For each assumed value of the precession rate we preform a least-squares fit to determine the best-fit values of the parameters $a, e$, and $\varpi_0$.\footnote{Note that for these fits each position estimate is given an equal weight because the uncertainties in the ringlet positions are dominated by systematic errors in the background subtraction and fitting procedure that are difficult to quantify {\it a priori}.} We then compute the $rms$ scatter of the residuals between the observed data and the best-fit model with that precession rate. This $rms$ statistic provides a measurement of how well the model fits the data, and should reach its minimum value when the assumed precession rate matches the real precession rate of this ringlet.

\begin{figure}[tbh]
\resizebox{6.5in}{!}{\includegraphics{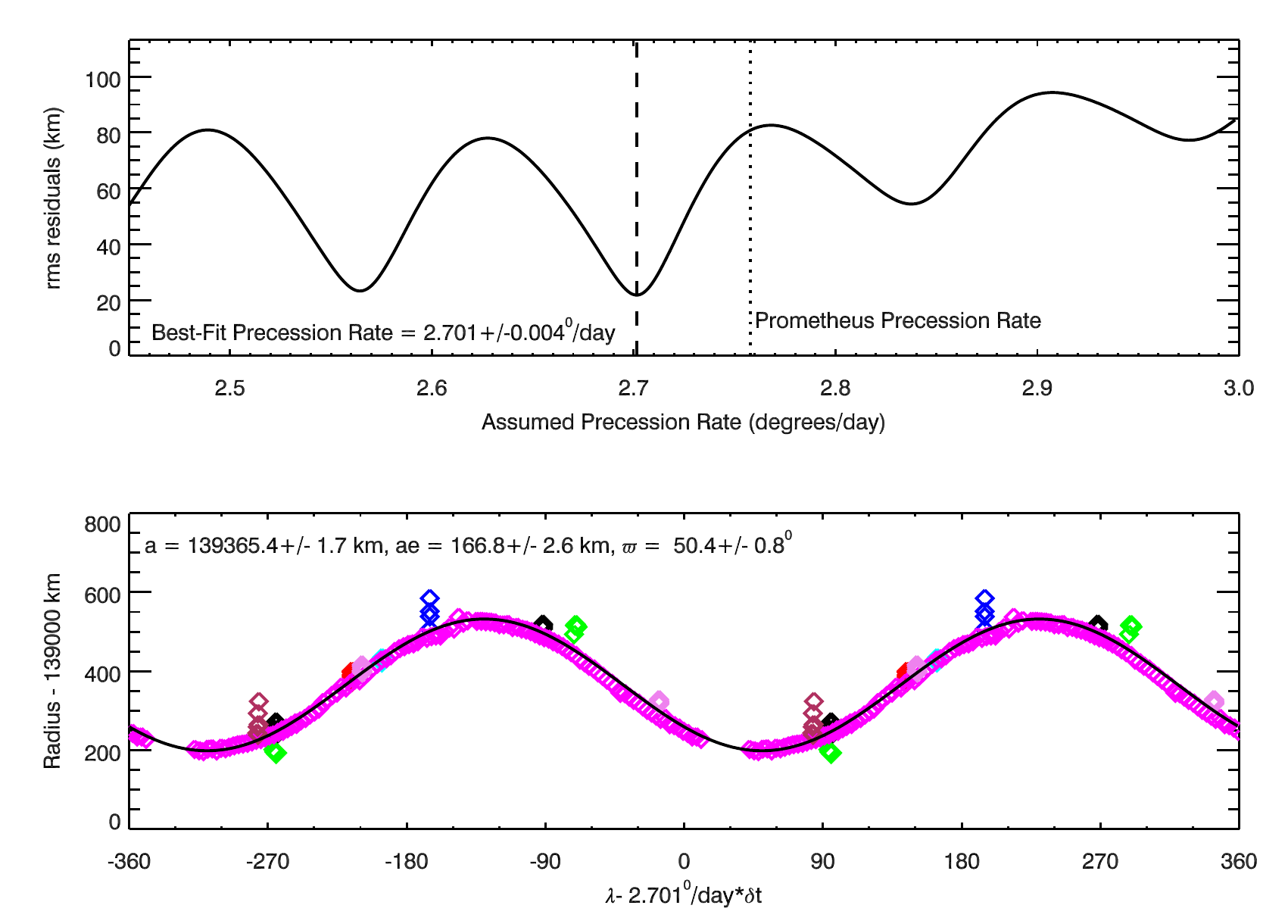}}
\caption{Same as Figure~\ref{fitalldat}, except that we only consider observations between 10$^\circ$ and 30$^\circ$ in front of Prometheus. This fit yields very similar parameters to those found with the entire dataset.}
\label{fitlondat}
\end{figure} 

\begin{figure}
\resizebox{6.5in}{!}{\includegraphics{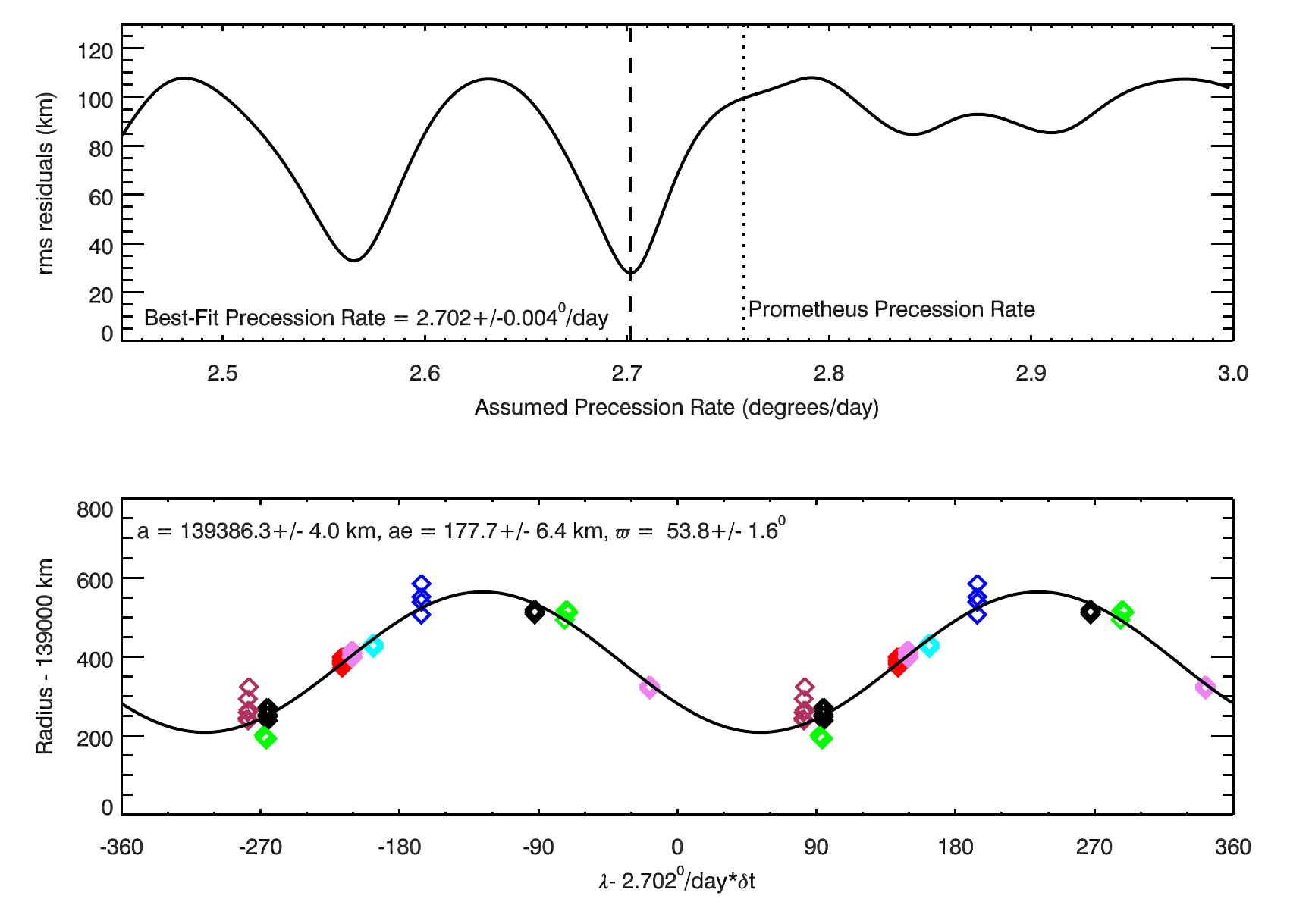}}
\caption{Same as Figure~\ref{fitalldat}, except that we only consider observations between 10$^\circ$ and 30$^\circ$ in front of Prometheus and exclude the FMOVIE199 data. This fit yields very similar parameters to those found with the entire dataset.}
\label{fitlondat2}
\end{figure} 

As shown in Figure~\ref{fitalldat}, the minimum in the $rms$ residuals occurs when the assumed precession rate is 2.702$\pm0.003^\circ$/day.\footnote{The uncertainty on this estimate derived from the shape of the $rms$ curve. Assuming that the minimum $rms$ corresponds to a reduced $\chi^2$ of 1, we estimate the probability to exceed as a function of pattern speed and fit this curve to a Gaussian. The Guassian width corresponds to the uncertainty in the precession rate.} This is a surprising result because the apsdial precession rate of Prometheus' orbit is 2.758$^\circ$/day \citep{Spitale06, Jacobson08}. This is also the expected precession rate for material with semi-major axes around 139,400 km given the current measurements of Saturn's gravity field \citep{Jacobson06}, so it would be natural to assume that the ringlet would precess at this rate. However, if we force the precession rate to match that of Prometheus' orbit, the data are clearly much less well organized (see Figure~\ref{fitpromdat}). This basic result also persists if we consider only the measurements taken between 10$^\circ$ and 30$^\circ$ in front of Prometheus (which should minimize radius variations with co-rotating longitude, see Figure~\ref{fitlondat}), and even if we eliminate the FMOVIE199 data from that data set (see Figure~\ref{fitlondat2}). Thus we are forced to conclude that despite the particles of this ringlet having a mean orbital radius that is close to Prometheus' orbit, this ringlet does not precess like Prometheus' orbit does. In fact, our best-fit precession rate is instead consistent with the precession rate of the F ring, which \citet{Albers12} found to be 2.70025$\pm0.00029^\circ$/day. The implications of this finding are discussed further below.

For the best-fit precession rate, we estimate the values and uncertainties on the other fit parameters $a$, $ae$ and $\varpi_0$ assuming the error on each data point is equivalent to the observed scatter in the residuals. Note that the $rms$ scatter of all the data around the best-fit model is roughly 40 km. By contrast, if we only consider data taken between 10$^\circ$ and 30$^\circ$ in front of Prometheus, the dispersion is closer to 30 km. This dispersion most likely represents a combination of fit uncertainties and real unmodelled variations in the ringlet's position. Since it is difficult to quantify these systematic uncertainties, we here use the $rms$ variations as a conservative estimate of the uncertainty in the position estimates and use these numbers to estimate the errors on the remaining fit parameters. The best-fit mean radius for the full data set is 139341.7$\pm$1.2 km and is   139365.4$\pm$1.7 km for the longitudinally-selected data set. The 23 km difference between these two estimates can be attributed to the variations in the ringlet's radial position with co-rotating longitude (see below). Note that both these estimates are slightly interior to Prometheus' semi-major axis of 139,380 km. By contrast, both data sets yield comparable estimates of the ringlet's $ae$, with the full data set giving $167.4\pm1.9$ km and the longitudinally-restricted data set giving $166.8\pm2.6$ km. Note that these numbers imply that the ringlet's eccentricity $e= 0.00120\pm0.00002$, which is about half of both Prometheus' orbital eccentricity of 0.0022 and the F-ring's eccentricity of 0.00235 \citep{Spitale06, Jacobson08, Albers12}. Finally, we may note that the pericenter position at the J2000 epoch time is about $50^\circ$ for both data sets, placing it about 25$^\circ$ in front of the F-ring's pericenter \citep{Albers12}, which is consistent with the ringlet's orientation in the FMOVIE199 observations (see Figure~\ref{fmovie199im}).
 
 \begin{figure}
\centerline{ \resizebox{3in}{!}{\includegraphics{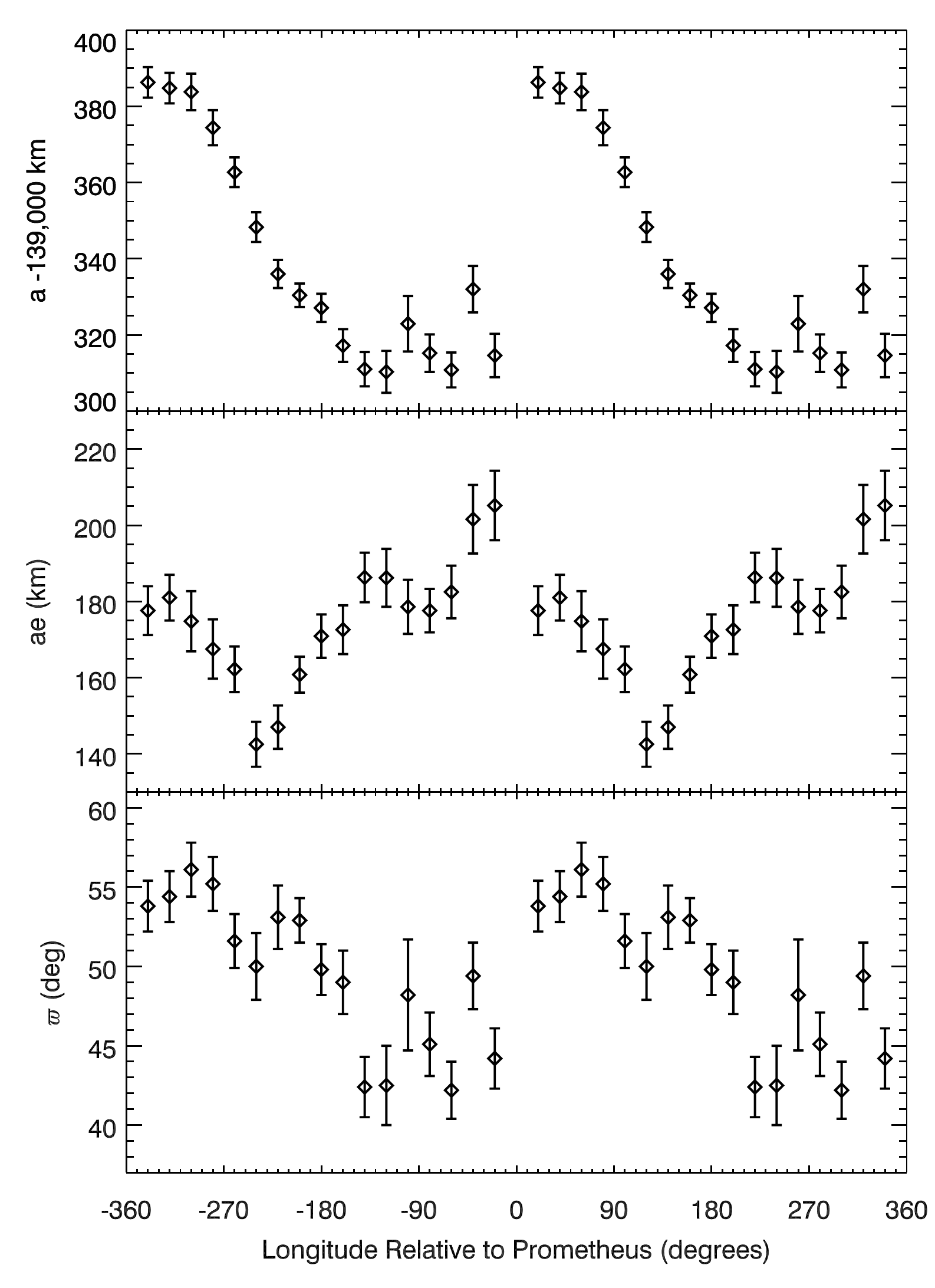}}}
 \caption{Ringlet parameters versus longitude relative to Prometheus, determined by fitting data in 20$^\circ$-wide bins and assuming a constant precession rate of 2.702$^\circ$/day. Note that the FMOVIE199 data are excluded in these fits, and that the data are repeated twice for clarity.}
 \label{params}
 \end{figure}
 
Finally, we examine how these orbital parameters vary with co-rotating longitude by fitting the data in 20$^\circ$-wide bins of longitude relative to Prometheus and assuming a fixed precession rate of 2.702$^\circ$/day. Figure~\ref{params} shows the resulting estimates of the ringlet's orbital parameters as functions of longitude relative to Prometheus (the fit parameters are also provided in Table~\ref{fittab}). Note the FMOVIE199 data are excluded from these fits in order to make the estimates more comparable, which means that the estimated mean radius and eccentricity are somewhat larger than the values given above (compare Figures~\ref{fitlondat} and~\ref{fitlondat2}). The most obvious trend in these data is in the ringlet's mean orbtial radius, which is close to Prometheus' semi-major axis of 139,380 km in front of Prometheus, but falls with increasing co-rotating longitude to about 139,320 km just behind Prometheus. By comparison, the variations in the eccentricity and pericenter locations are fairly modest. These may even reflect systematic errors among the various data sets rather than real structures in the ringlet.
 
\section{Discussion}
\label{discussion}

The most surprising aspect of the Prometheus ringlet's structure and dynamics is that  while its mean orbital radius appears to fall interior to Prometheus' orbit, its pericenter precesses at a rate close to that of the F ring, which is about $0.06^\circ$/day slower than one would expect for material close to Prometheus' orbit. While we do not yet have a model that can fully explain this phenomenon, we can highlight potentially relevant interactions with Prometheus and the F ring that are worthy of further investigation. 

In principle, the unusual behavior of the Prometheus ringlet could be explained in one of two ways: 
\begin{itemize}
\item[1.] The ringlet consists of material orbiting close to Prometheus' orbit, but some perturbation force acts on this material to slow the precession rate by about 0.06$^\circ$/day.
\item[2.] The ringlet actually consists of particles with semi-major axes close to the F ring whose eccentricities and pericenter locations are organized in such a way that they produce high concentrations of material close to Prometheus' orbit. 
\end{itemize}
While we cannot definitively rule out the second explanation, the observable structure of the ringlet does not favor that scenario. If the ringlet material had semi-major axes comparable to the F ring, then it should be drifting backwards relative to Prometheus and so should receive eccentricity kicks as it is passed by the moon. These kicks should give rise to periodic patterns analogous to the streamer-channel complexes in the F ring or the moonlet wakes in the A ring. However, such structures are not observed (see Figure~\ref{promimage}). 
 Instead the brightness of the ringlet varies slowly with distance from Prometheus, which is more consistent with material trapped in a 1:1 resonance with the moon. Furthermore, the mean radius of the Prometheus ringlet remains within about 70 km of Prometheus' semi-major axis. Material moving in the combined gravitational field of Prometheus and Saturn can maintain stable horseshoe motion around the moon's L3, L4 and L5 Lagrange points so long as their orbital semi-major axes are within a distance $\Delta a$ of Prometheus' semi-major axis $a_P$, where $\Delta a= f_h a_P(m_P/M_S)^{1/3}$, $m_P/M_S=2.8\times10^{-10}$ being the mass ratio between Prometheus and Saturn \citep{Jacobson08}, and $f_h$ being a numerical factor between 0.5 and 1.3 \citep{Weissman74, Dermott80, Goldreich82}. These numbers give $\Delta a$=45-130 km, so the Prometheus ringlet material is probably close enough to Prometheus' orbit to be trapped in a 1:1 co-rotation resonance, although it may approach the edge of the stability zone in the region just behind Prometheus. Hence for the remainder of this discussion we will assume that the ringlet material has semi-major axes close to that of Prometheus.

Assuming the ringlet particles do have roughly the same semi-major axes as Prometheus, then there needs to be some perturbation acting on these particles that is reducing their apsidal precession rates by $0.06^\circ$/day. The evolution of a particle's orbital elements in response to a generic perturbation force are given by the standard orbtial perturbation equations \citep{Burns76}. For material on nearly circular orbits, a perturbing force with a radial component $F_R$ and an azimuthal component $F_\lambda$ will cause the pericenter location to drift at a rate given by the following approximate expression \citep{Burns76}:
\begin{equation}
\frac{d\varpi}{dt}=\frac{n}{e}\left[-\frac{F_R}{F_G}\cos (\lambda-\varpi) +2\frac{F_\lambda}{F_G}\sin(\lambda-\varpi)\right]
\label{perieq}
\end{equation}
 where $n, e, \lambda$ and $\varpi$ are the particles' orbital mean motion, eccentricity, longitude and pericenter longitude, respectively, and $F_G=GM_P m_p/a^2$ is the central gravitational force from the planet on the particle ($M_P$ being the planet's mass, while $m_p$ is the particles' mass and $a$ is its semi-major axis). Given the ringlet's observed $e \simeq 0.0012$ and assuming the particles are orbiting at about the same rate as Prometheus (i.e. $n\simeq 587^\circ$/day), then changing the precession rate by $0.06^\circ$/day would require an orbit-averaged perturbation force of order $10^{-7}F_G$. 
  
 Since the Prometheus ringlet's precession rate is close to that of the F ring, the F ring is the most likely source of these perturbations. In principle. the perturbing force could either arise from the F-ring's finite mass density or be due to collisions between the ringlet particles and F-ring material. However, in practice the F-ring's gravity is unlikely to produce the observed reduction in the Prometheus-ringlet's precession rate. For one, the F-ring's gravity should have comparable effects on both Prometheus and the ringlet, and so it is difficult to imagine how this would cause the ringlet to precess at a different rate from Prometheus' orbit. Furthermore, the F-ring is probably not massive enough to produce such a large change in the precession rate. \citet{BGT83} calculated the precession rate that would be induced on a particle's orbit by a nearby ringlet. Following the \citet{NP88} notation, the relevant precession rate perturbation is:
 \begin{equation}
 \left.\frac{d\varpi}{dt}\right|_{grav}=\frac{m_F}{\pi M_P} n\left(\frac{a}{a-a_F}\right)^2F(q)\frac{e-e_F\cos(\delta \varpi)}{e_p}
 \end{equation}
 where $m_F$ is the F-ring's mass, $M_P$ is the planet's mass, $n, a$ and $e$ are the particles' mean motion, semi-major axis and orbital eccentricity, $a_F$ and $e_F$ are the F-ring's mean radius and eccentricity, $\delta \varpi$ is the difference in pericenter locations between the particles' orbit and the F ring, and the factor of $F(q)$ is:
 \begin{equation}
 F(q)=[(1-q^2)^{-1/2}-1]/q^2
 \end{equation}
 where
 \begin{equation}
 q^2=\left(\frac{a}{a-a_F}\right)^2\left[(e_F\sin\delta \varpi)^2+(e-e_F\cos\delta\varpi)^2\right]
 \end{equation}
Inserting numbers appropriate for the Prometheus ringlet and the F ring, we find that in order to reduce the precession rate of the particles in the Prometheus ringlet by 0.06$^\circ$/day, the mass of the F ring would need to be about fifty times larger than the mass of Prometheus, which is unreasonably large \citep{Murray96}. We therefore posit that the Prometheus ringlet is instead aligned with the F ring thanks to collisions with F-ring material.

A thorough analysis of such collisions would require numerical simulations and is beyond the scope of this paper, but we can provide some analytical arguments why this idea is reasonable. Consider the simple case of a single particle interacting with the material in a very narrow ringlet when both the particle and the ringlet orbit in the same plane. Say the particle has a semi-major axis $a$, eccentricity $e$ and pericenter location $\varpi_p$, and the ringlet has a mean radius equal to $a$, an eccentricity equal to $e$, but a different pericenter location $\varpi_r$. In this case, the particle will pass through the ringlet twice each orbit and exchange momentum with the ringlet material. \citet{Hedman10} stated that such collisions will tend to drive $\varpi_p$ towards $\varpi_r$, and here we will explicitly show this to be the case and thus demonstrate that inter particle collisions could potentially provide a mechanism for aligning eccentric tenuous ringlets.

So long as the eccentricities are small, the radial positions of the ringlet and the particle as functions of longitude $\lambda$ are given by the expression $r=a[1-e\cos(\lambda-\varpi)]$ and so the radial separation between the particle and the ringlet is:
\begin{equation}
\delta r =ae[\cos(\lambda-\varpi_p)-\cos(\lambda-\varpi_r)]=-2ae\sin\left(\frac{\varpi_p-\varpi_r}{2}\right)\sin\left(\frac{2\lambda-\varpi_p-\varpi_r}{2}\right)
\end{equation}
The particle will therefore cross the center of the ringlet at two longitudes $\lambda_c=(\varpi_p+\varpi_r)/2+\pi/2\pm\pi/2$. If the ringlet has a narrow radial width $W$ then the particle will be inside the ringlet over a range of longitudes:
\begin{equation}
\Delta \lambda \simeq \frac{W}{2ae|\sin[(\varpi_p-\varpi_r)/2]|}
\end{equation}
where this first-order approximation holds as long as $W<<2ae|\sin[(\varpi_p-\varpi_r)/2]|$ (for wider ringlets the full expression will asymtote to $\pi$).

In these regions, the material in the ringlet will be moving radially relative to the particle at a speed:
\begin{equation}
\delta v_r=e\sqrt{\frac{GM_P}{a}}[\sin(\lambda_c-\varpi_r)-\sin(\lambda_c-\varpi_p)]=\mp2e\sqrt{\frac{GM_P}{a}}\sin\left(\frac{\varpi_p-\varpi_r}{2}\right).
\end{equation}
(Note that the particle and the ringlet material have the same azimuthal speed, and we assume that all the ringlet material has the same radial speeds.) The particle will therefore feel perturbing forces in the radial direction as it passes through the ringlet and exchanges momentum with the ringlet material via collisions. A particle of radius $s_p$ moving through a ringlet at a relative speed $\delta v_r$ sweeps of a volume $\pi s_p^2 |\delta v_r|$ per unit time. If the ringlet has a local mass density $\rho_r$ then the momentum density of the ringlet in the particles' frame is $\rho_r\delta v_r$. The magnitude of the force applied to the particle while it is in the ringlet can therefore be expressed as:
\begin{equation}
|F_r|=\frac{\pi}{2} \rho_r s^2 \delta v_r^2={2\pi} \rho_r s_p^2 e^2\frac{GM_P}{a}\sin^2\left(\frac{\varpi_p-\varpi_r}{2}\right)
\label{forceeq}
\end{equation}
Note the factor of $1/2$ arises because the collisions with the ring material are not perfectly efficient at transferring momentum to the particle. Also note that this force is purely radial and can be inward or outward depending on the sign of $\delta v_r$.


 
 
 
Assuming that the ringlet force $F_r$ is applied when the particle is within the longitude range $\Delta \lambda$ of the two crossing longitude, then orbit-averaged precession rate induced in the particles' orbit by the ringlet:
 \begin{equation}
\left\langle \frac{d\varpi_p}{dt}\right\rangle=\frac{n}{e}\frac{W}{4\pi ae |\sin[(\varpi_p-\varpi_r)/2]|}\frac{|F_r|}{F_G}\left[\cos\left(\frac{\varpi_p+\varpi_r}{2}+\pi-\varpi_p\right)-\cos\left(\frac{\varpi_p+\varpi_r}{2}-\varpi_p\right) \right].
\end{equation}
Then, using Equation~\ref{forceeq} for $|F_r|$ and $GM_p m_p/a^2$ for $F_G$ (and taking care with absolute value signs) this expression becomes:
 \begin{equation}
\left\langle \frac{d\varpi_p}{dt}\right\rangle=-n\frac{\rho_r s_p^2 W}{m_p} \sin\left(\frac{\varpi_p-\varpi_r}{2}\right)\cos\left(\frac{\varpi_p-\varpi_r}{2}\right),
\end{equation} 
or, more simply:
 \begin{equation}
\left\langle \frac{d\varpi_p}{dt}\right\rangle=-n\frac{\rho_r s_p^2 W}{2m_p}\sin\left({\varpi_p-\varpi_r}\right).
\end{equation}  
This implies that any difference between the particles' pericenter and that of the ringlet should generate perturbation forces that drive the particles' pericenter to evolve in whatever direction will bring its pericenter into alignment with the ringlet, as desired. A similar result is obtained if one allows the ringlet and particle to have different semi-major axes. The only difference is that the applied force and induced precession rate go to zero at a nonzero value of $|\varpi_p-\varpi_r|$ because the particle's orbit no longer intersects the ringlet. 

In order to better ascertain the magnitude of this restoring force, it is useful to translate the ringlet's mass density $\rho_r$ into normal optical depth $\tau$. If we assume the ringlet consists of particles of size $s_r$ and mass $m_r$ with spatial density $\mathcal{N}$, then $\rho_r=\mathcal{N} m_r$ and so long as the optical depth is low enough, $\tau=\pi s_r^2 T\mathcal{N}$, where $T$ is the ringlet's vertical thickness. Thus we can re-write the induced precession rate as:
  \begin{equation}
\left\langle \frac{d\varpi_p}{dt}\right\rangle=-n \frac{\tau}{2\pi} \frac{W m_r s_p^2}{T m_p s_r^2}\sin\left({\varpi_p-\varpi_r}\right)
\end{equation}  
Assuming the ringlet has a radial width comparable to its vertical thickness, and that the particle is about the same size and mass of the ringlet material, this means the precession rate induced by collisions in the ringlet should be of order $\tau/2\pi$ time the particle's mean motion. Since the particles are orbiting at about the same rate as Prometheus, $n\simeq 587^\circ$/day, then perturbing the precession rate by $0.06^\circ$/day would require optical depths of order 0.0005. This implies that even fairly tenuous ringlets could potentially produce strong enough perturbing forces to align pericenters via collisions. While we do not yet have a direct measurement of the Prometheus ringlet's optical depth from occultation measurements, we can roughly estimate this parameters based on the ringlet's brightness. This ringlet is several hundred times fainter than the F ring \citep[which has a peak optical depth between 0.1 and 1, see][]{French14}, and so is probably also several hundred times lower in optical depth, which would imply a $\tau$ of order 0.001. The background Roche Division material outside the ringlet has higher brightnesses and optical depths, so it is reasonable to expect that interparticle collisions are indeed relevant to the dynamics of this system.

The above calculations support the idea that collisions between particles in the F ring and the Prometheus ringlet could be responsible for the latter's anomalous precession rate. Indeed, these arguments suggest that interparticle collisions could potentially play an important role in maintaining the structure of the entire F ring. However, more detailed analysis is needed to confirm this supposition. For example, the Prometheus ringlet never intersects the core of the F ring, so the alignment between these two rings needs to be mediated by material orbiting between them. Hence the real system is much more complex, involving a large number of particles with a range of orbital semi-major axes. Most likely, the ability of particle collisions to align orbital pericenters in these situations will depend upon on the radial distribution of material, and all these particles will probably only precess at the same rate as the F-ring core if the core has sufficient mass and/or optical depth. Furthermore, these collisions should not only perturb the particles' precession rates, but could also influence their orbital eccentricities, and one might even hope that a complete model of this system would yield the low but non-zero eccentiricity of the observed Prometheus ringlet. Investigating these topics will most likely require numerical simulations that are well beyond the scope of this report, but hopefully, future research along these lines will be able to explain the Prometheus ringlet's precession rate, as well as its observed eccentricity and orientation relative to the F ring. 

Further studies of particle collisions within the Prometheus ringlet may also clarify the origin of the trends in the ringlet's mean orbital radius with longitude relative to Prometheus shown in Figure~\ref{params}. As mentioned above, these variations are reminiscent of trends found in the orbital parameters of the Central Encke Gap Ringlet, which consists of material trapped in a 1:1 co-rotation resonance with Pan. The mean radius of the Encke Gap ringlet systematically increases behind Pan from 0 to 10 km exterior to Pan's orbital semi-major axis \citep{Hedman13}. These variations have been interpreted as evidence that the small ringlet particles are spiraling outwards under the influence of plasma drag. Such outward migration would cause the particles to have a slower mean motion than Pan, and so the material drifts both backwards and outwards relative to that moon to produce the observed trends in that ringlet's location. The observed trends in the Prometheus ringlet's mean radius could be explained using a very similar model, but it is important to note that while the Encke Gap ringlet is found {\em exterior} to Pan's orbit, the Prometheus ringlet is found {\em interior} to Prometheus' orbit. This suggests that the ring material would be moving inwards, causing it to drift forwards relative to Prometheus. This inward migration cannot be due to simple plasma drag because the Prometheus ringlet lies in a region where the magnetospheric plasma rotates around the planet faster than the Keplerian rate, and so momentum exchange with the plasma would naturally cause the particles' semi-major axes to increase, not decrease. Thus some other perturbing force must be responsible for the mean radius variations in the Prometheus ringlet. In principle, collisions with F-ring material could induce inwards migration, but the efficiency of this process is sensitive to such phenomena as the collision geometry and how dissipative the collisions are. Thus more detailed analysis will be needed to ascertain whether interactions with the F ring are responsible for the observed longitudinal trends in the Prometheus ringlet's location.

\begin{table}
\caption{Shape parameters for the Prometheus ringlet}
\label{fittab}
\centerline{\begin{tabular}{|c|c|c|c|c|}\hline
Fit & $\dot{\varpi}$ & $a$ & $ae$ & $\varpi_0^a$ \\ 
 & (degrees/day) & (km) & (km) & (degrees) \\ \hline
All data & 
2.703$\pm$0.003 & 139341.7$\pm$1.2 & 167.4$\pm$1.9 & 50.1$\pm$0.6 \\
All data with $10^\circ<\lambda-\lambda_P<30^\circ$ & 
2.701$\pm$0.004 & 139365.4$\pm$1.7 & 166.8$\pm$2.6 & 50.4$\pm$0.8 \\
Data with $10^\circ<\lambda-\lambda_P<30^\circ$, & & & & \\
except for FMOVIE199 & 
2.702$\pm$0.004 & 139386.3$\pm$4.0 & 177.7$\pm$6.4 & 53.8$\pm$1.6 \\
\hline
Data with $10^\circ<\lambda-\lambda_P<30^\circ$, & & & & \\
except for FMOVIE199 & 
2.702 (fixed) & 139386.3$\pm$4.0 & 177.7$\pm$6.4 & 53.8$\pm$1.6 \\
Data with $30^\circ<\lambda-\lambda_P<50^\circ$ & 
2.702 (fixed) & 139384.8$\pm$4.0 & 181.0$\pm$6.0 & 54.4$\pm$1.6 \\
Data with $50^\circ<\lambda-\lambda_P<70^\circ$ & 
2.702 (fixed) & 139383.8$\pm$4.8 & 174.8$\pm$7.9 & 56.1$\pm$1.7 \\
Data with $70^\circ<\lambda-\lambda_P<90^\circ$ & 
2.702 (fixed) & 139374.4$\pm$4.6 & 167.5$\pm$7.8 & 55.2$\pm$1.7 \\
Data with $90^\circ<\lambda-\lambda_P<110^\circ$ & 
2.702 (fixed) & 139362.7$\pm$3.9 & 162.2$\pm$6.0 & 51.6$\pm$1.7 \\
Data with $110^\circ<\lambda-\lambda_P<130^\circ$ & 
2.702 (fixed) & 139348.3$\pm$3.9 & 142.5$\pm$5.9 & 50.0$\pm$2.1 \\
Data with $130^\circ<\lambda-\lambda_P<150^\circ$ & 
2.702 (fixed) & 139336.0$\pm$3.7 & 147.0$\pm$5.7 & 53.1$\pm$2.0 \\
Data with $150^\circ<\lambda-\lambda_P<170^\circ$ & 
2.702 (fixed) & 139330.4$\pm$3.1 & 160.8$\pm$4.7 & 52.9$\pm$1.4 \\
Data with $170^\circ<\lambda-\lambda_P<190^\circ$ & 
2.702 (fixed) & 139327.1$\pm$3.7 & 170.9$\pm$5.7 & 49.8$\pm$1.6 \\
Data with $190^\circ<\lambda-\lambda_P<210^\circ$ & 
2.702 (fixed) & 139317.2$\pm$4.3 & 172.6$\pm$6.4 & 49.0$\pm$2.0 \\
Data with $210^\circ<\lambda-\lambda_P<230^\circ$ & 
2.702 (fixed) & 139311.0$\pm$4.5 & 186.3$\pm$6.5 & 42.4$\pm$1.9 \\
Data with $230^\circ<\lambda-\lambda_P<250^\circ$ & 
2.702 (fixed) & 139310.3$\pm$5.5 & 186.2$\pm$7.6 & 42.5$\pm$2.5 \\
Data with $250^\circ<\lambda-\lambda_P<270^\circ$ & 
2.702 (fixed) & 139322.9$\pm$7.3 & 178.6$\pm$7.1 & 48.2$\pm$3.5 \\
Data with $270^\circ<\lambda-\lambda_P<290^\circ$ & 
2.702 (fixed) & 139315.2$\pm$4.9 & 177.6$\pm$5.7 & 45.1$\pm$2.0 \\
Data with $290^\circ<\lambda-\lambda_P<310^\circ$ & 
2.702 (fixed) & 139310.8$\pm$4.6 & 182.5$\pm$6.9 & 42.2$\pm$1.8 \\
Data with $310^\circ<\lambda-\lambda_P<330^\circ$ & 
2.702 (fixed) & 139332.0$\pm$6.1 & 201.6$\pm$9.0 & 49.4$\pm$2.1 \\
Data with $330^\circ<\lambda-\lambda_P<350^\circ$ & 
2.702 (fixed) & 139314.6$\pm$5.7 & 205.2$\pm$9.1 & 44.2$\pm$1.9 \\ \hline
\end{tabular}}

$^a$ Pericenter position at J2000 epoch relative to ascending node of Saturn's ringplane on J2000.

\end{table}

\section*{Achknowledgements}

This work was supported by NASA Cassini Data Analysis Program Grant NNX15AQ67G. We thank the imaging team, the Cassini project and NASA for the data used for this analysis. We also thank P.D. Nicholson, C.D. Murray, D.P. Hamilton, M.R. Showalter and J.A. Burns for useful conversations. We thank Robert French and Sebastien Charnoz for their helpful reviews of this manuscript.

\section*{Supplemental Information}

A machine-readable table of all our estimates of the Prometheus-ringlet's radial position, along with the corresponding observation times and longitudes, are provided as supplemental online information to this manuscript.

 \end{document}